\documentclass[acmtog,authorversion,nonacm]{acmart}
\acmJournal{TOG}

\title{High-Order Continuous Geometrical Validity}

\author{Federico Sichetti}
\affiliation{%
\institution{University of Genoa}
\country{Italy}
}
\email{federico.sichetti@edu.unige.it}

\author{Zizhou Huang}
\affiliation{%
\institution{New York University}
\country{United States}
}
\email{zizhou@nyu.edu}

\author{Marco Attene}
\affiliation{%
\institution{CNR-IMATI: GENOVA}
\country{Italy}
}
\email{marco.attene@ge.imati.cnr.it}

\author{Denis Zorin}
\affiliation{%
\institution{New York University}
\country{United States}
}
\email{dzorin@cs.nyu.edu}

\author{Enrico Puppo}
\affiliation{%
\institution{University of Genoa}
\country{Italy}
}
\email{enrico.puppo@unige.it}

\author{Daniele Panozzo}
\affiliation{%
\institution{New York University}
\country{United States}
}
\email{panozzo@nyu.edu}

\usepackage[english]{babel}
\usepackage{libertinus}
\usepackage{algpseudocodex}
\usepackage{algorithm}
\usepackage{amsfonts}
\usepackage{amsmath}
\usepackage{amsthm}
\usepackage{mathtools}
\usepackage{enumerate}
\usepackage{csquotes}
\usepackage{hyperref}
\usepackage{xcolor}
\usepackage{multirow}
\usepackage{tikz-3dplot}
\usepackage{siunitx}
\usepackage{arydshln} %
\usepackage{wrapfig}
\usepackage[capitalize,noabbrev]{cleveref}

\newcommand{\True}{\textsc{true}}
\newcommand{\False}{\textsc{false}}

\newcommand{\reals}{\mathbb{R}}
\newcommand{\ints}{\mathbb{Z}}
\newcommand{\nats}{\mathbb{N}}

\newcommand{\intervals}{\mathbb{I}}
\newcommand{\lagpoly}{\mathcal{L}}
\newcommand{\bezpoly}{\mathcal{B}}

\newcommand{\timedep}[1]{\bar{#1}}
\newcommand{\transition}[2]{\mathbf{T}_{#1\to#2}}

\newcommand{\indexset}{\mathcal{I}}
\newcommand{\cornerset}{\mathcal{C}}
\newcommand{\domainpts}{\Gamma}
\newcommand{\intlo}[1]{\underline{#1}}
\newcommand{\inthi}[1]{\overline{#1}}
\newcommand{\inclusion}[1]{\Box #1}
\newcommand{\minclusion}[1]{\Box_{\min} #1}
\newcommand{\sampleset}[1]{\mathcal{D}_{#1}}

\newcommand{\Continue}[1]{\textbf{continue}}
\newcommand{\Break}[1]{\textbf{break}}

\newtheorem{definition}{Definition}

\DeclareMathOperator*{\argmin}{arg\,min}

\begin{document}

\begin{teaserfigure}
    \centering
    \includegraphics[width=\linewidth]{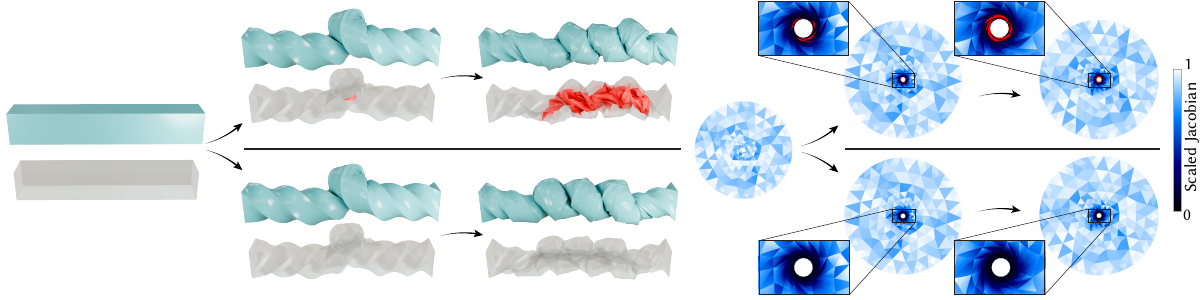}
    \caption{Soft body simulations \cite{highorderIPC} of twisting a beam in 3D (left) and twisting the inner circle of a fat ring in 2D (right), without (top) and with (bottom) our continuous validity check. The high-order invalid elements are marked in red, and the scaled Jacobian distribution is shown on the right. Without our method, the Twist-beam example has 1 and 161 invalid elements at t=\qty{3}{\s} and t=\qty{7}{\s} respectively; the Twist-ring example has 11 and 31 invalid elements at t=\qty{0.4}{\s} and t=\qty{1}{\s} respectively. With our validity check and adaptive quadrature refinement scheme, the simulations do not contain any invalid elements.}
    \label{fig:teaser}
\end{teaserfigure}

\begin{abstract}

We propose a conservative algorithm to test the geometrical validity of simplicial (triangles, tetrahedra), tensor product (quadrilaterals, hexahedra), and mixed (prisms) elements of arbitrary polynomial order as they deform within a time interval.

Our algorithm uses a combination of adaptive Bézier refinement and bisection search to determine if, when, and where the Jacobian determinant of an element's polynomial geometric map becomes negative in the transition from one configuration to another.
In elastodynamic simulation, our algorithm guarantees that the system remains physically valid during the entire trajectory, not only at discrete time steps. 
Unlike previous approaches, physical validity is preserved even when our method is implemented using floating point arithmetic. Hence, our algorithm is only slightly slower than existing inexact methods while providing guarantees and while being an easy drop-in replacement for current validity tests.

To prove the practical effectiveness of our algorithm, we demonstrate its use in a high-order Incremental Potential Contact (IPC) elastodynamic simulator and experimentally show that it prevents invalid, simulation-breaking configurations that would otherwise occur using inexact methods.
    
\end{abstract}

\maketitle

\section{Introduction}
In computer graphics, mechanical engineering, and scientific computing, physical objects are often modeled using meshes composed of simple elements, such as tetrahedra, hexahedra, and prisms. 
Each element is typically associated with two maps: (1) a geometric map %
that defines the element's shape; and (2) a basis map that extends quantities (such as displacement or velocities) defined at the element's nodes into its interior. For rendering, linear polynomials (hat functions) are commonly used for both maps, but higher-order versions are widely employed when greater accuracy is required.
When the basis map is used to interpolate a \emph{displacement}, %
the element's geometry is derived by combining the initial geometric map with the basis map, yielding a polynomial whose order corresponds to the highest of the two. 
A typical example in graphics is the use of second-order elements in finite element (FE) simulations for fabrication \cite{Panetta2015}, which generates quadratic curved elements even if the initial geometric map is piecewise linear.

\begin{figure*}
    \centering	\includegraphics[width=0.9\linewidth]{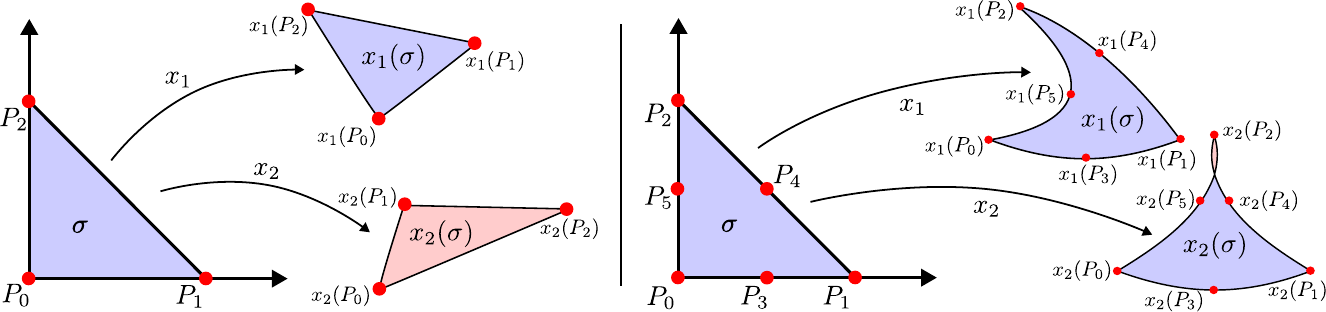}
	\caption{A reference domain $\sigma$ is mapped to linear elements (Left) and quadratic elements (Right).  
    Quadratic elements can be given either by an initial quadratic geometry or by combining an initial linear geometry with a quadratic displacement.
 Geometric maps on a Lagrangian basis %
    are specified by mapping the domain points $P_i$ to control points $x_j(P_i)$. %
    Blue and red areas denote a positive and negative determinant of the Jacobian, respectively. 
    In both %
    cases, %
    element $x_2(\sigma)$ is invalid. 
    In the linear case, the whole element is inverted, while in the quadratic case, only a small portion of $x_2(\sigma)$ is inverted. 
    }
    \label{fig:canonical}
\end{figure*}

In the following, we will overload the term \emph{geometry map} to describe the final geometry of an object, which can thus also incorporate the displacement basis map.
In mathematical notation, this is defined with a polynomial function:
$$x:\sigma\longrightarrow\reals^n,$$
where $\sigma$ represents a reference element domain (such as a standard simplex or multi-interval), and $n$ denotes the dimension of the element. 
See \cref{fig:canonical} for examples.

During animation or simulation, objects are deformed by modifying the geometry map, the basis map, or both: these modifications are usually performed %
by applying linear transformations to the map coefficients. 
Note that, for a piecewise linear geometric map, this reduces to the usual interpolation of the coordinates of the mesh vertices. 
Similarly, for curved meshing, it is also typical to start with a piecewise linear mesh and then curve its elements to reduce the approximation error \cite{Toulorge2013}. 

Since these meshes are used to represent physical objects, a basic, yet elusive, requirement is that the geometry of the object does not self-intersect at any time during deformation.
This condition can be violated in two ways: (1) by a global intersection or (2) by a local lack of injectivity of the geometric map. 
The first problem has been studied thoroughly in the parametrization and simulation literature, at least for linear elements \cite{Smith:2015,Bolun:2021}. 
The second condition has been considered mostly in the context of mesh parametrization, and, even in that case, no robust algorithm exists (see Section \ref{sec:sota}). 
In this work, we focus on the latter.

\begin{figure}
	\centering
	\begin{tikzpicture}[scale=.4]
	\newcommand{\mytriangle}[6]{
		\coordinate[label=#4: $A$] (A) at (0,0);
		\coordinate[label=#5: $B$] (B) at #1;
		\coordinate[label=#6: $C$] (C) at #2;

		\draw[thick, fill=#3] (A) -- (B) -- (C) -- cycle;

		\foreach \point in {A, B, C}
        \fill[red] (\point) circle (5pt);
	}
	\newcommand{\mytrianglearr}[6]{
		\mytriangle{#1}{#2}{#3}{#4}{#5}{#6};
		\draw[->] (B) -- +(-1,-1);
		\draw[->] (C) -- +(0,-2);
	}
	\begin{scope}[shift={(0,4)}]
		\mytrianglearr{(5,1)}{(2,3)}{blue!15}{left}{right}{above};
		\node[rectangle, draw=black] (id) at (2,-1) {$t=0.0$};
	\end{scope}
	\begin{scope}[shift={(8,5.5)}]
		\mytrianglearr{(4,0)}{(2,1)}{blue!20}{left}{right}{above};
		\node[rectangle, draw=black] (id) at (2,-2.5) {$t=0.2$};
	\end{scope}
	\begin{scope}[shift={(15,7)}]
		\mytrianglearr{(3,-1)}{(2,-1)}{red!20}{left}{right}{below left};
		\node[rectangle, draw=black] (id) at (3,-4) {$t=0.4$};
	\end{scope}
	\begin{scope}[shift={(0,0)}]
		\mytrianglearr{(2,-2)}{(2,-3)}{red!20}{above left}{above right}{right};
		\node[rectangle, draw=black] (id) at (2,1) {$t=0.6$};
	\end{scope}
	\begin{scope}[shift={(8,1)}]
		\mytrianglearr{(1,-3)}{(2,-5)}{blue!20}{above}{left}{right};
		\node[rectangle, draw=black] (id) at (2,0) {$t=0.8$};
	\end{scope}
	\begin{scope}[shift={(16,1)}]
		\mytriangle{(0,-4)}{(2,-7)}{blue!20}{above}{left}{right};
		\node[rectangle, draw=black] (id) at (2,0) {$t=1.0$};
	\end{scope}
\end{tikzpicture}
	\caption{A dynamic linear element with linear trajectories that flips twice in a time interval, resulting in an element that is valid at both time steps, but invalid in the transition. Assuming point $A$ remains fixed, the arrows represent the velocities of points $B$ and $C$ to reach the final position. The color of each element indicates the sign of the determinant at time $t$: blue if positive and red if negative. The dynamic element's Jacobian determinant is a univariate polynomial in $t$ of degree $2$, with two distinct roots in $[0,1]$. }
	\label{fig:fliptwice1}
\end{figure}
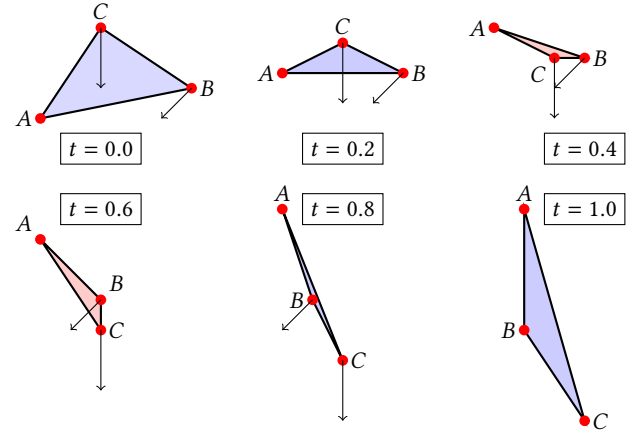

\paragraph{Problem statement}
We denote with $\timedep{x}$ a geometric map that is dynamically changing over time and we study its injectivity. 
This reduces to studying the sign of the determinant of its Jacobian $|J_{\timedep{x}}|$ to assess the continuous validity of the element, i.e., if,  where and when $|J_{\timedep{x}}|$ becomes negative. 
We note that this is a subtly challenging problem \emph{even} for the simple case of a linear triangle (\cref{fig:fliptwice1}): an element that is valid in its initial and final configurations might become invalid \emph{during} the trajectory. 

\paragraph{Brief summary of the state of the art} 
A common approach among practitioners is testing the validity of elements by just computing the value of $|J_x|$ at quadrature points at every time step, e.g., \cite{Dey2001}. 
This might fail to detect invalidity even in static tests, let alone the continuous case. 
\citet{Smith:2015} study the continuous problem for the specific case of 2D linear triangles, reducing it to a quadratic root finding, which is solved numerically; this 
approach is hard to extend to 3D or to higher-order polynomials.
Other approaches exist to resolve the static test for high-order elements, e.g., \cite{johnen_geometrical_2014}, but they are neither conservative nor easily extensible to the continuous test. 
More detail on the state of the art is provided in \cref{sec:sota}.

\paragraph{Contribution}
We introduce the first generic formulation and algorithm for the continuous validity test of elements, supporting the most common types — such as triangles, quadrilaterals, tetrahedra, prisms, and hexahedra — and extending to high-order basis and geometric maps. Our algorithm is provably conservative when implemented using floating-point arithmetic, meaning that if an element is detected to be valid, it is guaranteed to remain valid throughout the entire specified time interval. This level of robustness, crucial for algorithmic reliability, has not been achieved by any previous method.

If an element becomes invalid at any point, our algorithm provides a conservative estimate of the inversion time and introduces a custom quadrature rule that accurately reflects the detected inversion. Specifically, this means that the numerical integration diverges when the element inverts, a property not provided by adaptive quadrature rules commonly used in high-order finite elements.

While designed for the dynamic case, our algorithm can also be used for the static case, with minor modifications: in this setting, our algorithm is the first algorithm to provide a conservative static geometrical validity test for high-order elements.

\paragraph{Evaluation}

Our algorithm is designed and implemented for high performance, as its use-case is within optimization loops requiring the testing of large datasets: on static checks, we demonstrate that our test is competitive in terms of runtime with current inexact methods, being slightly slower while guaranteeing a conservative answer.
To quantitatively evaluate the correctness and efficiency of our approach and compare it with more specialized alternatives, we construct a dataset of 2D and 3D time-dependent queries whose ground truth is computed using (extremely expensive) symbolic root finding.

\paragraph{Applications}
Having access to a conservative check we discovered that it is very common for high-order FE simulations to contain invalid elements in their solution; we show examples in PolyFEM in \cref{fig:teaser,fig:simulation-flip}.
This seems to be a common problem with high-order FE codes: for example, FEBio \cite{Maas2012} also uses a static check only at quadrature points.
This issue is rarely mentioned in the literature \cite{Anderson2014,Dobrev2019} and we are not aware of other papers proposing a solution.
We believe that the presence of invalidity is due to the use of insufficiently accurate quadrature to capture the infinite elastic potential inside some of the most distorted elements. This is a major source of both the numerical fragility of this software and inaccuracy in the solution as physically invalid configurations are reported as the simulation result. By replacing the validity check and the quadrature in PolyFEM with our approach, we show that these issues disappear and the impact on performance is moderate.

\paragraph{Impact}
We believe our algorithm will be an essential addition to the growing toolkit of robust geometric building blocks used in modern parametrization, meshing, and simulation algorithms. To foster its adoption, we will release an open-source reference implementation.

\begin{figure*}
	\centering
	\includegraphics[width=\linewidth]{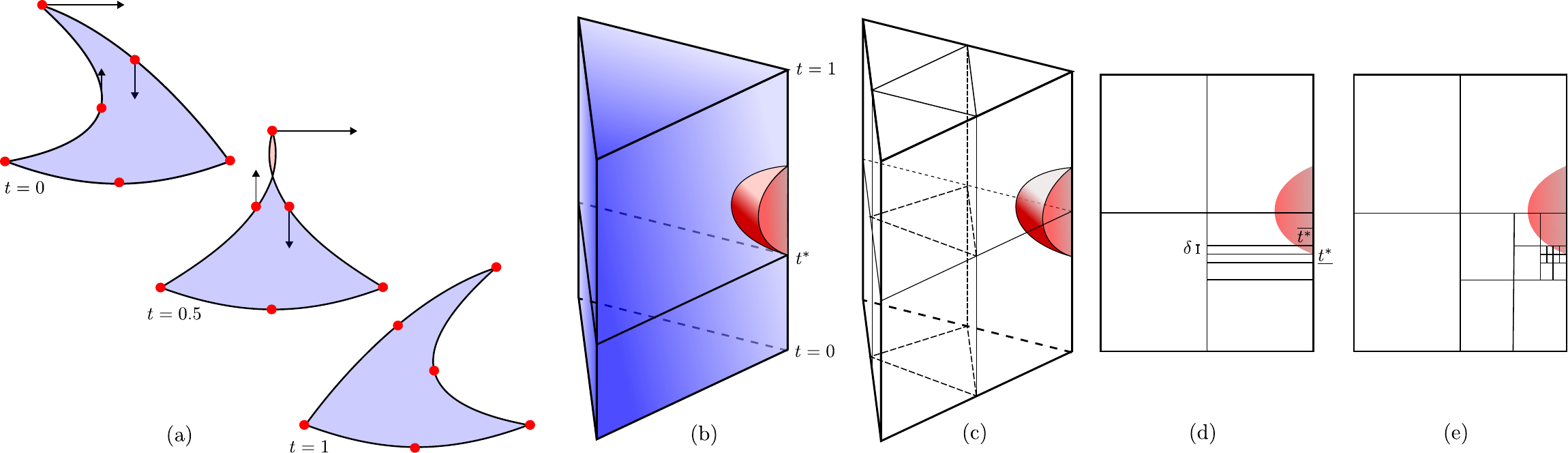}
	\caption{(a): A 2D quadratic element with linear trajectories is valid at the start and end positions but locally inverted during the transition. 
(b): Its parametric domain is an upright triangular prism, where the vertical axis represents time and each horizontal slice represents the space domain at a given time. 
 The red region denotes the portion of the domain in which the determinant $|J_{\timedep{x}}|$ is not positive.
 The dynamic element's Jacobian determinant is a trivariate polynomial in $\xi_1, \xi_2, t$ of order 2 in its space variables and order 2 in the time variable.
 (c): The root finding algorithm bisects the time domain and quadrisects the spatial domain to isolate a thin slice $[\intlo{t^*},\intlo{t^*}+\delta]$ containing the critical time of inversion $t^*$. 
 In the first step, both space and time dimensions are refined, generating eight sub-domains, which are pushed into a priority queue. 
 (d): For clarity we show a side view for the following steps. Only the time dimension is bisected: the analysis of sub-domains that do not intersect the invalid region increases the value of $\underline{t^*}$ while the analysis of sub-domains that do intersect the invalid region decreases the value of $\overline{t^*}$ until convergence. 
 (e): If all dimensions were refined at all steps, many more sub-domains would be generated, thus killing performance. 
 }
	\label{fig:running-example}
\end{figure*}

\section{Overview of the method}
Given a dynamic element $\timedep{x}(\sigma)$ deforming over the time interval $[0,1]$, we %
study 
the determinant of its Jacobian $|J_{\timedep{x}}|$ 
(\cref{sec:formulation,sec:method}). 
The polynomial $|J_{\timedep{x}}|$ can have a high order in both its spatial and temporal variables (\cref{app:bases}). 
To ensure a conservative answer, we employ a custom bisection root-finding method controlled by an accuracy parameter $\delta > 0$. Let $t^*$ denote the earliest time at which $\timedep{x}(\sigma)$ becomes invalid (i.e., $|J_{\timedep{x}}|$ turns negative). 
We return a time $\intlo{t^*} > t^* - \delta$ and a point $\intlo{P} \in \sigma$ such that $|J_{\timedep{x}}|$ is positive everywhere for $t\leq\underline{t^*}$, and $|J_{\timedep{x}}(\intlo{P},t)|$ becomes negative for some $t \leq \intlo{t^*} + \delta$.
This guarantees that the element can safely deform up to time $\intlo{t^*}$ while the point $\intlo{P}$ is used to adaptively refine the quadrature rule, steering the simulation away from invalid configurations (\cref{sec:applications}). 
If the element remains valid throughout, we simply return $\intlo{t^*} = 1$.

The parameter domain of a dynamic element has both space and time dimensions. 
Our test proceeds by bisecting the time dimension: it maintains lower $\underline{t^*}$ and upper $\overline{t^*}$ bounds for the critical inversion time, and terminates when the difference between these bounds is smaller than %
$\delta$, or when the element is confirmed to be valid throughout the interval. 
The algorithm employs a priority queue of sub-domains, which are created by recursively splitting the initial space-time domain. %

For a given sub-domain $S$, we compute a \emph{minimum inclusion function} that returns an interval $I$, which is guaranteed to contain the minimum value of $|J_{\timedep{x}}|(S)$: if $I$ is strictly positive, the element is valid in $S$; if $I$ is strictly negative, the element is invalid somewhere within $S$ (but it is not necessarily invalid everywhere in $S$); if $I$ contains zero, nothing can be said and further refinement of $S$ is necessary. 
The minimum inclusion function is a crucial component of our method; details on its definition and computation are provided in \cref{sub:inclusion,sub:inclusion-ST}, respectively.

Another key aspect of the algorithm is the decoupling of refinements in the spatial and temporal dimensions. 
While the time dimension may require refinement until the interval between $\underline{t^*}$ and $\overline{t^*}$ is lower than $\delta$, the spatial dimensions usually need less refinement. 
In essence, bisection along the time axis is primarily driven by the need to narrow the bounds of $t^*$, while subdivision of the spatial domain is employed to resolve indeterminate configurations. %
Details and pseudo-code are given in \cref{sub:algorithm}.

To account for numerical errors, we employ interval arithmetic.
The value of parameter $\delta$ allows for a trade-off between computation time and the accuracy of the estimation; however, regardless of the parameter choice, the algorithm always provides a conservative estimation.

\paragraph{Example}
Consider the quadratic element in \cref{fig:running-example}(a). 
It is valid at $t=0$, becomes invalid at some intermediate time $t^*<0.5$, and then returns to a valid state at a later time, remaining valid until $t=1$. 

The parameter domain for this dynamic element can be visualized as in \cref{fig:running-example}(b) with an upright triangular prism, where each horizontal slice of the prism represents the spatial domain at a specific time, and time progresses along the vertical axis from 0 to 1. Notably, the element inversion occurs within a localized wedge (red volume in the figure) during the times when the element is invalid. 
In general, inversions can occur anywhere within the domain, including regions away from vertices.

The minimum inclusion function is evaluated first for the whole prism. 
The result is an interval that contains zero, thus the domain is subdivided as in \cref{fig:running-example}(c) along both the time and the space dimensions, and eight sub-prisms are pushed to the priority queue.
While processing the four prisms corresponding to the time interval $[0,0.5]$, %
the minimum inclusion function returns a strictly positive interval for three of them, which are discarded from the queue; and it returns a strictly negative interval for the fourth one, which intersects the red wedge: this domain is bisected only in the time dimension and its two children are pushed onto the queue. 
Sub-domains spanning earlier times are processed first.

In the subsequent refinements (\cref{fig:running-example}(d), side view), the minimum inclusion function will always be strictly negative, hence only bisection in the time dimension will occur: the analysis of valid intervals that do not intersect the red wedge contributes to increasing the value of $\underline{t^*}$ while the analysis of intervals that do intersect the red wedge contributes to decreasing the value of $\overline{t^*}$, until convergence.
Note that, if the domain were always subdivided along space and time, as in \cref{fig:running-example}(e), many more subdomains would be generated and the algorithm could become much slower.
For the 3D order 3 armadillo dataset, with 10\% target error, the naive approach has an average processing time of $\sim\!\!260\mu s$ per element, whereas our approach takes $\sim\!\!14\mu s$ per element on average. For higher element orders, or queries with higher precision, the maximum number of subdivisions used in the naive approach must be limited (otherwise the check will consume unreasonable amounts of memory), and the test can fail to produce an estimate within precision on some elements.

\section{Related work}\label{sec:sota}

We briefly describe polynomial bases and their use in graphics, methods for checking static and continuous validity of elements, and discuss their application in FEA and meshing. We conclude with an overview of robust predicate evaluation techniques, which we use in our algorithm.

\paragraph{High-Order Bases and Geometry for Finite Element Analysis}

Linear basis functions are often a sub-optimal choice in many contexts. For example, \citet{Schneider2022} advocate for the use of high-order bases for elliptic PDEs; \citet{Bargteil2014}, \citet{Mezger2009}, \citet{Suwelack2013}, \citet{Ushakova2011} use high-order elements for animation; and \citet{Mandad2020} propose to use a high-order basis for parametrization. 
Our algorithm provides a guaranteed-conservative check for the validity of these elements, increasing the robustness of any method using a high-order basis.

A related, but distinct, concept is the use of high-order geometry, where the geometry of an element is represented using a high-order polynomial. More commonly, $C^0$  \cite{Jiang2021} or $C^k$ geometric maps (IGA) \cite{Cottrell2009} are used. The latter option is popular in mechanical engineering, where IGA envisions the use of the same representation for both interpolating the physical quantities and for representing the geometry.

Despite the different uses, our algorithm applies to generic polynomials, hence it can be used as is to ensure the validity of the geometric map.

\paragraph{Static Element Inversion Check}

The special case of checking the geometric validity of a linear triangle/tetrahedron in a static setting has been solved in a seminal paper by \citet{richard_shewchuk_adaptive_1997}, where a robust predicate called \textsc{orient2d}/\textsc{orient3d} is introduced. This paper revolutionized mesh generation and simulation, providing a reliable, yet efficient, solution to one of the basic primitives used by meshing and simulation algorithms. To the best of our knowledge, this approach has not been extended to elements with higher order; such extension is challenging because an element may flip at certain points while remaining valid at others, as in the example shown on the right side of \cref{fig:canonical}.

For high-order simplicial elements, a common approach to test for element inversion consists in testing only their quadrature points \cite{polyfem,Maas2012,gargallo-peiro_distortion_2015,Dey2001}: while effective at avoiding NaNs in the integration of certain diverging elastic potentials, this approach is not conservative, leading to incorrect stresses (\cref{fig:nan-stress}). 
An efficient but not conservative method has been introduced in \cite{johnen2013,johnen_geometrical_2014,JOHNEN2018}, where the Jacobian determinant of the element is represented in Bézier form: the inversion check then reduces to testing the positivity of the Bézier coefficients as they undergo adaptive refinement. This method can miss inversions (we provide a numerical example in \cref{app:gmsh}) due to the use of floating point arithmetic. We take inspiration from this check and similarly use adaptive Bézier subdivision to derive a continuous test: our algorithm solves a different problem by adding the time dimension and is designed to be conservative due to its judicious use of rational and interval arithmetic.

For the special case of hexahedral elements, which are commonly used in commercial finite element analysis software, this problem has been extensively studied in \cite{Ushakova2011}. Unfortunately, these tests are insufficient to guarantee validity but are used nonetheless due to their efficiency. \citet{Vavasis_2003} proposes a sufficient condition but does not provide a conservative algorithm that takes advantage of it. \citet{Johnen_Weill_Remacle_2017} propose an optimized version of \cite{johnen_geometrical_2014} specifically for linear hexahedral elements, however, the approach still suffers from the same floating point issues as the original test.

A radically different approach is taken in \cite{marschner_hexahedral_2020}, where a Sum-of-Squares (SOS) relaxation is used to compute the minimum Jacobian determinant, reducing the problem to solving a sequence of small semidefinite programming problems of increasing complexity.   The method offers generality and an elegant formulation, but is relatively expensive computationally (multiple SDP solves per element), and only guarantees injectivity up to numerical precision of an iterative convex solver,  which might result in invalid elements.

All the methods described above are designed to handle static checks exclusively, which is insufficient for validating deforming elements over time. 

\paragraph{Continuous Element Inversion Check}

To the best of our knowledge, the only paper explicitly addressing the problem of checking the validity of an element that deforms over time is \cite{Smith:2015}. They propose an algorithm to estimate the safest step before an inversion for 2D linear elements by using the closed form of the roots of a degree 2 polynomial. 
Their approach is however not conservative due to floating point rounding, as we show in \cref{sec:results}. Furthermore, the method does not scale to linear 3D elements -- where the polynomial is of order 3 and robust root finding is not trivial -- nor higher order elements -- where the polynomial is multivariate and closed forms for the roots may not even be available. 

The problem is also discussed in \cite{Anderson2014} for high-order element remapping and in \cite{Dobrev2019} for high-order meshing, however no algorithm for the validity check is proposed. Our algorithm could be used in their setting to provide a conservative validation of their resulting elements.

\paragraph{High-Order Meshing}
High-order meshing requires a high-order validity check to ensure that elements are valid after curving. We refer to \cite{Geuzaine2015,jiang_bijective_2021} for an overview of the state of the art of high-order meshing. Our contribution can be used, in its reduced form for static validity, as a provably conservative check in any of these meshing algorithms to increase their robustness.%

\paragraph{Interval Arithmetic}

According to the IEEE 754 standard, the result of a floating point operation is a rounded value of the exact result. Instead of computing a single rounded value, one can compute an interval that \emph{contains} the exact result by simply rounding in both directions. Replacing FP numbers with intervals slows the calculations by an order of magnitude on average but, on the other hand, enables a provably correct implementation of geometric algorithms \cite{snyder92}. Intervals are provided by many existing libraries such as CGAL \cite{bronnimann98}, Boost \cite{boost}, Filib and Filib++ \cite{filib}, BIAS \cite{bias} and GAOL \cite{gaol}. We point the reader to \cite{benchmark_interval} for a detailed comparison of these tools. When %
full portability is not required, modern SIMD architectures can be exploited to speed up interval arithmetic significantly. The basic idea is to store both the bounds in a single register that can host 128 bits, which is the space required by two double precision FP numbers, and then perform each operation on the entire register simultaneously. This approach, first introduced in \cite{lambov}, is employed in the numeric kernel of the indirect predicates library \cite{ipred_lib}, which we use in our implementation. 

We use intervals to represent inclusion functions and to guarantee that any possible rounding error is always tracked so as to provide a correct polynomial evaluation.

\section{Preliminaries and notations}
\label{sec:prelim}

We address $n$-dimensional ($n=2, 3$) high-order meshes consisting of elements of various types. 
The geometry of every element is defined by a polynomial map. 
We will refer to the \emph{order} $p$ of a polynomial as the maximum exponent of a single variable
(as opposed to the usual notion of degree).
\cref{tab:syms} gives a summary of symbols defined in the following and used throughout.

\begin{table}
	\begin{tabular}{c|l|c}
		Symbol & Meaning & Def.\\
		\hline
            $n$ & dimension of element and embedding space & \ref{sec:prelim}\\
            $s$ & dimension of simplicial part of element & \ref{sub:mix}\\
            $\sigma, \sigma^n_s$ & static reference element (with dimensions) & \ref{sub:mix}\\
            $\xi_1,\ldots\xi_n$ & spatial coordinates & \ref{sub:mix}\\
            $p$ & order of an element / polynomial & \ref{sub:maps}\\
            $x$ & geometric map of an element& \ref{sub:maps}\\
            $|J_x|$ & Jacobian determinant of $x$ & \ref{sub:maps}\\
            $\intervals$ & intervals on the real line &   \ref{sub:inclusion}\\
            $\inclusion{f}$ & inclusion function for $f$ & \ref{sub:inclusion}\\
            $\minclusion{f}$ & minimum inclusion function for $f$ & \ref{sub:inclusion}\\
            $t$ & time coordinate & \ref{sec:formulation}\\
            $\timedep{\sigma}, \timedep{\sigma}^n_s$ & dynamic reference element (with dim.) & \ref{sec:formulation}\\
            $\timedep{x}$ & dynamic geometric map of an element & \ref{sec:formulation}\\
            $t^*$ & minimum time at which $|J_{\timedep{x}}|$ vanishes & \ref{sec:formulation}\\
            $\intlo{t^*}, \inthi{t^*}$ & lower and upper bounds to $t^*$ & \ref{sec:formulation}\\
            $\delta$ & user-specified accuracy & \ref{sec:formulation}\\
            $l_{\max}$ & maximum level of recursion & \ref{sec:formulation}\\
            $\psi^-, \psi^+$ & time-only subdivision maps & \ref{sec:formulation}\\
            $\psi^q$ & $q$-th subdivision map & \ref{sec:formulation}\\
            $\cornerset_\sigma^p$ & subset of indices of corners of element $\sigma$ & \ref{sec:bezier} \\
            $\transition{\bezpoly}{\bezpoly}^\pm$ & time-only subdivision matrix & \ref{sec:bezier}\\
            $\transition{\bezpoly}{\bezpoly}^q$ & $q$-th subdivision matrix & \ref{sec:bezier}\\
		$\domainpts_{\sigma}^p$ & set of domain points of order $p$ on $\sigma$ & \ref{app:bases}\\
            $\gamma_i$ & domain point & \ref{app:bases}\\
            $\indexset_\sigma^p$ & set of indices of points of $\domainpts_{\sigma}^p$ & \ref{app:bases}\\
            $\lagpoly^m_{i}$ & $i$-th Lagrange polynomial of order $m$ & \ref{app:bases}\\
            $\bezpoly^m_{i}$ & $i$-th Bernstein polynomial of order $m$ & \ref{app:bases}\\
            $f^\lagpoly$ & vector of coefficients in Lagrange form & \ref{app:bases}\\
            $f^\bezpoly$ & vector of coefficients in Bézier form & \ref{app:bases}\\
            $\transition{\lagpoly}{\bezpoly}$ & transition matrix from Lagrange to Bézier & \ref{app:bases}\\
	\end{tabular}
	\caption{Symbols used in the text and where they are defined. }
	\label{tab:syms}
\end{table}

\subsection{Reference Element Domains}
\label{sub:mix}
For each type of element, we define a common \emph{reference domain} (or \emph{reference element}) $\sigma^n_s\subset[0,1]^n$ to use as the coordinate domain.

\begin{definition}[Reference Domain]
\label{def:element}
	Let $n, s \in \nats$, $1 \leq s \leq n$. The $n$-dimensional reference domain $\sigma^n_s\subset[0,1]^n$ is the locus of points with coordinates $(\xi_1,\ldots,\xi_n)$ that satisfy the system of inequalities:
	\begin{gather}
		\nonumber \xi_i \geq 0 \quad \forall i\in\{1, \dots, n\} \\
		1 - \sum_{i=1}^s \xi_i \geq 0 \\
		\nonumber \xi_i \leq 1 \quad \forall i\in\{n - s + 1, \dots, n\} \label{eq:ref_ineq}
	\end{gather}
\end{definition}

With this notation, we have a general parameter space that works for all the most commonly used FEM elements. 
The element $\sigma^1_1$ is the unit segment; $\sigma^2_1$ is the unit square; $\sigma^2_2$ is the standard triangle; $\sigma^3_1$ is the unit cube; $\sigma^3_2$ is the unit triangular prism; $\sigma^3_3$ is the standard tetrahedron.
More generally, $\sigma^n_s$ is the tensor product of a standard $s$-simplex with a standard $(n-s)$-hypercube\footnote{Note that, the unit segment can be seen both as a 1-simplex and as a 1-hypercube; to avoid any ambiguity, we always treat it as a simplex, so that, e.g., the unit square is regarded as the tensor product of a 1-simplex with a 1-hypercube, rather than as a 2-hypercube.}.

\begin{figure}
	\centering
	\includegraphics[width=\columnwidth]{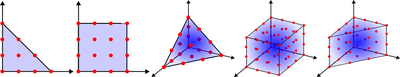}
	\caption{Reference elements with domain points of order 3: triangle, square, tetrahedron, hexahedron, prism.}
	\label{fig:domains}
\end{figure}

\begin{figure}
\centering	\includegraphics[width=0.8\columnwidth]{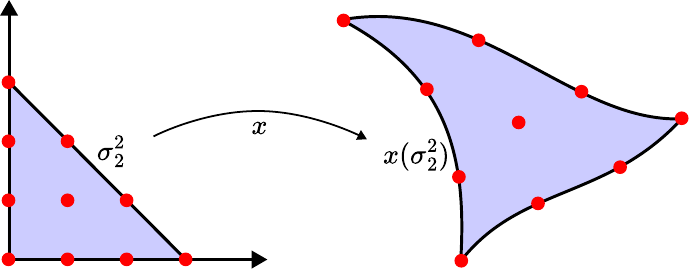}
 \vspace{3mm}	\includegraphics[width=0.8\columnwidth]{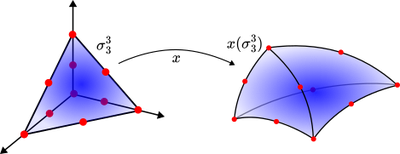}
	\caption{Above: A domain $\sigma^2_2$ with its domain points %
 of order 3 and the geometric map $x$ to the physical element $x(\sigma^2_2)$ with Lagrange control points. Below: likewise for an element $\sigma^3_3$ of order 2.  }
\label{fig:geometric-map}
\end{figure}

\subsection{High-Order Elements}
\label{sub:maps}
A generic polynomial $f:\sigma\longrightarrow\reals$ can be defined with a basis of polynomials: 
we consider here the Lagrange basis -- which is most common in FEM -- and the Bernstein basis that gives the Bézier form.
Both representations define $f$ as a linear combination of the basis functions with \emph{control coefficients} %
associated with \emph{domain points} that form a regular grid over the reference element; the number of domain points sets the \emph{order} $p$ of the polynomial (\cref{fig:domains}). 

For a given order and dimension, pre-computed conversion matrices allow us to convert between these two representations in both directions. 
The Lagrange and Bézier forms and the conversion matrices are detailed in \cref{app:bases}.

\paragraph{Geometric Map}
The geometric map 
$x:\sigma^n_s\longrightarrow\reals^n$ that maps a reference element $\sigma^n_s$ into the physical element $x(\sigma^n_s)$ is represented by specifying its set of control points, where each control point is a $n$-dimensional point (\cref{fig:geometric-map}). 
Each coordinate of $x$ is expressed with a multivariate polynomial, which is the tensor product of a $s$-variate polynomial of degree $p$ with $(n-s)$ univariate polynomials of degree $p$, each in a different variable.
It follows that the map $x$ has 
order $p$ in all its variables (even though its total degree may be higher).

In the following, we study the determinant of the Jacobian $J_x$ of the geometric map $x$, denoted $|J_x|$, which is also a multivariate polynomial in the same variables as $x$, but of a different order.
In particular, the number of terms of $|J_x|$ rapidly increases with the dimension and order of the element.
See 
\cref{app:bases} for details.

\subsection{Minimum Inclusion Function}
\label{sub:inclusion}
Informally speaking, given a real function $f$ and a domain $D$, an \emph{inclusion function} for $f$ over $D$ returns an interval that bounds the range of values of $f$ in $D$. 
Inclusion functions are widely used in root finding and in the evaluation of robust predicates \cite{snyder92}.
In our case, we are rather interested in an interval that just contains the \emph{minimum} value of $f$.

Let $\intervals$ %
be the space of intervals on the real line. 
For $a=[\intlo{a},\inthi{a}]\in\intervals$, let us define $w(a)=\inthi{a}-\intlo{a}$ the \emph{width} of interval $a$.
Let $A=a_1\times\cdots\times a_n\in\intervals^n$ be a $n$-dimensional interval; 
we extend the definition of width as $w(A)=\max_{j=1}^n w(a_j).$
Given $D\subseteq\reals^n$ compact, we further extend the definition of width as $w(D)=\min_{A\supseteq D} w(A)$.

Let $\Omega\subseteq\reals^n$ be a compact domain, let $f$ be a real function defined on $\Omega$, and let us denote $\mathcal{P}(\Omega)$ the subsets of $\Omega$.
Given a function $f:\Omega\longrightarrow\reals$, an \emph{inclusion function} for $f$ is a function 
$\inclusion{f}:\mathcal{P}\Omega \longrightarrow \intervals$ such that, for any $D\subseteq\Omega$ we have 
$$\forall \xi\in D \ \ \ f(\xi)\in\inclusion{f}(D).$$
We say $\inclusion{f}$ to be \emph{convergent} if for any $D\subseteq\Omega$ 
$$w(D)\rightarrow 0 \Rightarrow w(\inclusion{f}(D))\rightarrow 0.$$
In particular, if $D$ shrinks about $\xi$, then $\inclusion{f}(D)$ shrinks about $f(\xi)$.
\begin{definition}[Minimum inclusion function]
\label{def:minclusion}
A minimum inclusion function for the function $f$ is a function  
$\minclusion{f}:\mathcal{P}(\Omega) \longrightarrow \intervals$ such that, for any $D\subseteq\Omega$ we have 
$$\min_{\xi\in D} f(\xi)\in\minclusion{f}(D).$$
\end{definition}
If the lower end of $\minclusion{f}(D)$ is positive, we know that $f$ is always positive in $D$; if the upper end is negative, we know that $f$ has negative values, but is not necessarily negative everywhere, in $D$; otherwise, nothing can be said about the sign of $f$ in $D$.

A convergent inclusion function can be used to find a root of a function $f$ by subdividing the initial domain $\Omega$ until it becomes sufficiently small.
Likewise, one 
can use a convergent minimum inclusion function to find the portions of $\Omega$ where $f$ is positive, by recursively subdividing the domain.
The type of subdivision used to perform refinement depends on the shape of $\Omega$.
For instance, while bisection can be used for a multi-interval domain, simplicial domains may require less trivial subdivision rules (\cref{app:subdivision}). 

\subsection{Interval and Rational Arithmetic}
\label{sub:interval}
Interval arithmetic consists of a set of operations defined on the set of real intervals $\intervals$ such that if $\xi\in A_{\xi} \in \intervals$ and $\zeta\in A_{\zeta} \in \intervals$,
then $(\xi*\zeta) \in A_{\xi} * A_{\zeta}$, where $*$ in the right-hand side is the interval version of the operation.
If the exact result of an operation on floating point numbers falls between two representable values, rounding is required. 
We make our computations conservative by replacing floating point numbers with singleton intervals (i.e., intervals with matching endpoints), and rounding the left end of the result down and the right end up for every subsequent interval operation.
In our code, we use the implementation of \citet{attene20}.
More details are provided in \cref{app:interval}.

Interval arithmetic does not prevent the propagation of error; it merely keeps track of it. 
One way to implement exact arithmetic is via rational numbers: as long as an algorithm only involves rational operations, we can represent numbers exactly as fractions of integer values since every floating point number is also rational.
The main drawback of rational arithmetic is that it can be orders of magnitude slower than interval arithmetic because the bits needed to encode each fraction increase with computations.
For this reason, our use of rational arithmetic is limited to off-line computations.

\section{Continuous geometrical validity}
\label{sec:formulation}

In a dynamic simulation, the elements of the mesh move and deform over time. 
Like space, time is discretized into %
time steps, which are typically regular.
We assume that the control points move along straight-line trajectories at each time step, and, without loss of generality, we can assume each transition occurs between time $t=0$ and time $t=1$. Following \cref{def:element} we have:

\begin{definition}[Dynamic reference element]
	Let $\sigma^n_s$ be a reference element. The dynamic element $\timedep{\sigma}^n_s$ of $\sigma^n_s$ is the $(n+1)$-dimensional reference element $\sigma^{n+1}_s = \sigma^n_s\times[0,1]$.
 \label{def:dynamic}
\end{definition}

Assuming linear trajectories,
the \emph{dynamic geometric map} 
$\timedep{x}:\reals^{n+1}\to\reals^{n}$
of order $p$ for $\timedep{\sigma}^n_s$ is expressed by linear interpolation 
of the $n$-dimensional geometric maps $x^0(\xi)$ and $x^1(\xi)$ of the static element at the two consecutive time steps:
\begin{equation}
 \timedep{x}(\xi,t) = x^0(\xi) + t(x^1(\xi)-x^0(\xi)).
\label{eq:time-dependent}
\end{equation}
This map is of order $p$ in the $\xi$ variables and linear in $t$.

\subsection{Continuous Validity Test}
\label{sub:algorithm}
Assuming that the input element is valid at time $t=0$, the \emph{continuous validity test} consists of determining whether or not the Jacobian determinant  $|J_{\timedep{x}}(\xi)|$ is everywhere greater than 0 on $\bar{\sigma}$ at all times in $[0,1]$.
If this is not the case, the algorithm should find the earliest time $t^*\in(0,1]$ in which the element becomes invalid -- hence, it is valid within $[0,t^*)$.

\paragraph{Target accuracy}
Since finding an exact solution is not necessary and very expensive, we instead settle for a conservative estimate $\intlo{t^*}$ that is close to $t^*$ up to a given user-provided threshold $\delta$, i.e. $t^*\in[\intlo{t^*}, \intlo{t^*} + \delta]$. Moreover, since we assumed that $t^*>0$, we require that $t^*$ be strictly positive as well, regardless of the value of $\delta$.
This threshold is 
used to trade off accuracy and time performance. 

\paragraph{Early termination}
An intrinsic challenge of both the static and dynamic problems is that certifying the positivity of a polynomial can be arbitrarily hard. Therefore, there is no upper bound on the number of subdivisions required to assess the validity of an element. 
 
To prevent the algorithm from taking unreasonable time, we employ a termination criterion that triggers when refinement becomes excessive.
In these extreme cases, we halt and provide an estimate $\intlo{t^*}$ that may not be within $\delta$ of the true value $t^*$, but is nevertheless conservative.

Let the \emph{depth} of a subdomain $S$ be the number of subdivisions required to obtain $S$ from the initial domain.
In our implementation we stop when the depth of a subdomain $S$ exceeds a threshold $l_{\max}$.

\paragraph{Minimum inclusion function.} 
We rely on a convergent inclusion function, which comes together with a procedure to decompose an element into sub-elements. 
For the sake of clarity and generality, we first describe our algorithm avoiding the details on how we define our inclusion function and domain subdivision strategy, which we detail in \cref{sec:bezier}.
We also use $J=|J_{\timedep{x}}|$ as a short-hand notation for the determinant of the Jacobian of the dynamic element at hand.

Given a generic inclusion function $\inclusion J$ on the domain $\timedep{\sigma}^n_s$, we define a minimum inclusion function as follows: let $\inclusion J(D)=[\intlo{J_D},\inthi{J_D}]$ then
$$\minclusion J(D)=[\intlo{J_D},\min_{\timedep{\xi_i}\in\sampleset{D}}J(\timedep{\xi_i})],$$
where $\sampleset{D}$ is a small set of samples in $D$.
In practice, we sample $J$ at these points to bound the minimum of $J$ from above.
Note that, if $J$ is negative at any of those samples, we know that the element becomes invalid in $D$.

\paragraph{Subdivision Maps}
Given a reference domain $\sigma=\sigma^n_s$, we define a set of $Q$ linear maps $\{\psi^q: \sigma \longrightarrow \sigma\}_q$ called the \emph{subdivision maps} of $\sigma$, and we call $\psi^q(\sigma)$ a \emph{subdomain} of $\sigma$; we require that the union of all subdomains is $\sigma$, and the intersection of any two subdomains is either empty or has dimension less than $n$.
We define the standard subdivision maps for $n\in\{2,3\}$ that we use in our implementation in \cref{app:subdivision}. In the following, we always assume that $Q = 2^n$.

For a time dependent reference domain $\timedep{\sigma}=\timedep{\sigma}^n_s$, its subdivision maps are the same as $\sigma^{n+1}_s$, but we also define two additional \emph{time subdivision maps}, denoted $\psi^-$ and $\psi^+$, such that $\psi^-(\timedep{\sigma})$ and $\psi^+(\timedep{\sigma})$ are respectively the lower and upper half of $\timedep{\sigma}$ when bisected in the time dimension only.

\begin{algorithm}[htb]
 \caption{Maximum valid time step with inclusion functions}\label{alg:bezierDynamic}
	\begin{algorithmic}[1]

	\Function{MaxValidStep}{$J, \delta, l_{\max}$}
		\State $P\gets$\Call{PriorityQueue}{$\prec$} \label{alg1:pq}
		\Comment{priority queue for subdomains}
		\State $\overline{t^*} \gets 1$ \Comment{initialize upper bound of $t^*$}
		\State $\underline{t^*} \gets 0$ \Comment{initialize lower bound of $t^*$}

		\State \Call{Push}{$P, \sigma$} \label{alg1:pushed}

		\State $F \gets \False$
        \Comment{flag of whether an invalidity has been found}
		\State $l \gets 0$
		\Comment{maximum subdivision depth reached so far}
		\While{\True}
			\If{$F \wedge (\overline{t^*}-\underline{t^*}\leq\delta) \wedge (\underline{t^*} > 0)$}
			\Comment{reached accuracy} \label{alg1:precision}
    			\State \Return $\intlo{t^*}$
                \Comment{conservative estimate of $t^*$}
			\EndIf

            \If{\Call{IsEmpty}{$P$}}
            \State \Return 1
            \EndIf

			\State $S\gets$\Call{Pop}{$P$} \label{alg1:pop}
			\Comment{get the next subdomain from $P$}

			\State $l\gets \max\{l,$ \Call{Depth}{$S$}$\}$
			\Comment{update maximum depth}
			\If{$l>l_{\max}$} \label{alg1:maximum}
				\Comment{maximum level reached: give up}
			\State \Return $\intlo{t^*}$
            \Comment{conservative estimate of $t^*$}
			\EndIf

			\State $\underline{t^*} \gets$ \Call{StartTime}{$S$}
			\Comment{everything before this time is valid}

			\State $I\gets$\Call{$\minclusion{J}$}{$S$}
			\Comment{check minimum inclusion}

			\If{\Call{High}{$I$} $\leq 0$} \label{alg1:negative}
				\Comment{there is an invalidity in $S$}
				\If{\Call{EndTime}{$S$}$<\overline{t^*}$}
		              \State $F \gets \True$
					\State $\overline{t^*}\gets$ \Call{EndTime}{$S$}
				\EndIf
				\State \Call{Push}{$P,\psi^-(S)$}\Comment{bisect on the $t$ axis only}
				\State \Call{Push}{$P,\psi^+(S)$}\Comment{bisect on the $t$ axis only}

			\ElsIf{$\lnot($\Call{Low}{$I$} $> 0)$} \label{alg1:fullsplit}
				\For{$q \in \{1,\dots,Q\}$}
					\State \Call{Push}{$P,\psi^q(S)$}\Comment{subdivide on $\xi$ and bisect on $t$}
				\EndFor
			\EndIf
			\EndWhile
	\EndFunction

	\Function{$\prec$}{$S_0, S_1$} \label{alg1:prec}
		\Comment{priority function}
		\If{\Call{StartTime}{$S_0$} $\neq$ \Call{StartTime}{$S_1$}}
			\Comment{lower time first}
			\State \Return \Call{StartTime}{$S_0$}$<$\Call{StartTime}{$S_1$}
		\Else
			\Comment{for ties, prioritize boxes most likely to be invalid}
			\State \Return \Call{High}{$\minclusion{J}(S_0)$}$<$\Call{High}{$\minclusion{J}(S_1)$}
		\EndIf
	\EndFunction

	\end{algorithmic}
\end{algorithm}

\paragraph{Pseudo-Code} The pseudo-code of the algorithm is given in \cref{alg:bezierDynamic}.
The algorithm takes in input a polynomial $J(\xi,t)$ defined on domain $\timedep{\sigma}$ and the thresholds $\delta$ and $l_{max}$ and returns a time $\intlo{t^*}$ without invalid configurations, and within a time $\delta$ of an invalid configuration.
The algorithm keeps internal current lower and upper bounds for $t^*$, initializing them to 0 and 1, respectively. 
Given a subdomain $S\subset \timedep{\sigma}$, pseudocode functions $\Call{StartTime}{S}$ and $\Call{EndTime}{S}$ return respectively the minimum and maximum values of the time coordinate for points in $S$.

The algorithm uses a priority queue $P$ (line \ref{alg1:pq}) of subdomains of $\sigma$, with a related priority function $\prec$ (line \ref{alg1:prec}) giving higher priority to sub-domains that span intervals of time with an earlier start point (that is, a lower minimum $t$).
The initial domain $\sigma$ is pushed into the priority queue $P$ (line \ref{alg1:pushed}); then elements are popped from the $P$ one by one (line \ref{alg1:pop}), and if their minimum inclusion function does not guarantee their validity, they are subdivided and their subdomains are pushed to $P$. 
See the next paragraph for the subdivision strategy.
This continues until the queue becomes empty or an early exit condition is met.
By construction, the priority function $\prec$ guarantees that when we pop an element $S$ from the queue, then $J$ is positive at all times before $\Call{StartTime}{S}$, and $\intlo{t^*}$ can then be updated accordingly.

Early exits occur if the required accuracy $\delta$ is achieved (line \ref{alg1:precision}), meaning that the difference between $\inthi{t^*}$ and $\intlo{t^*}$ is less than $\delta$, or the maximum depth $l_{\max}$ has been reached (line \ref{alg1:maximum}), meaning that we pop from the queue an interval that comes from a sequence of $l_{\max}$ subdivisions.

\paragraph{Subdivision Strategy}
If the interval $I$ returned by $\minclusion{J}(S)$ is completely negative -- meaning that $S$ contains negative values of $J$ (line \ref{alg1:negative}) -- then the upper bound $\inthi{t^*}$ is updated, and element $S$ is bisected along the time dimension only, with the two resulting subdomains being pushed into $P$.
Conversely, if interval $I$ contains zero, the element $S$ is split along all its dimensions (line \ref{alg1:fullsplit}), including time, according to the subdivision scheme of reference element $\timedep{\sigma}$; again, the resulting elements are pushed into $P$.
Finally, no subdivision is necessary if $I$ only contains positive values, and the space-time region occupied by $S$ will not be considered again for the remainder of the algorithm.

Note that, by bisecting only the time dimension (line \ref{alg1:negative}), we postpone any refinement of the spatial dimensions until we find a time interval in which $J$ may potentially be positive everywhere. 
This strategy allows us to avoid many unnecessary refinements in the space dimensions, and to decouple the subdivision on time (controlled by accuracy $\delta$) from the subdivision in space (which does not have an accuracy requirement).

\section{Implementation}
\label{sec:bezier}
\label{sec:method}

The implementation of our method requires designing a minimum inclusion function $\minclusion f$ and a corresponding subdivision strategy that uses robust computations while keeping the runtime sufficiently low to enable its use within a simulation loop.

\subsection{Inclusion Functions for Space and Time}
\label{sub:inclusion-ST}
As observed by \citet{snyder92}, interval arithmetic provides a universal way to design inclusion functions. 
Given a polynomial $f:\Omega\longrightarrow\reals$ and $D\subseteq\Omega$, let $A_{D}$ be the smallest multi-interval containing $D$. 
We could define
$$\inclusion f(D)=f(A_{D}),$$
where the evaluation of $f$ on the right side is intended with interval arithmetic, and thus returns an interval. 
Any strategy subdividing $D$ and reducing $A_D$ (e.g., bisection along all coordinates) provides a convergent inclusion function.

We tried this approach, but the inclusion functions may be very loose about $f$ and require many refinement steps to converge, or even get stuck on nearly invalid elements due to the numerical error accumulating too fast for the inclusion function to keep up with.
We compare the time performance for the static case only in \cref{tab:static}.
We instead follow the approach proposed by \citet{johnen_geometrical_2014} for the static validity test and extend it to our continuous setting.

\paragraph{Overview of Bézier Refinement}

Let $f$ be the order $m$ polynomial of which we want to find the minimum on $\timedep{\sigma}$ (in our case, $f=|J_{\timedep x}|$). 

Our inclusion function is based on the Bézier representation of $f$ and a recursive decomposition of $\timedep{\sigma}$. 
The reason why we want to represent our polynomial in the Bézier basis is the convex hull property \cite{Farin:2001}, by which the values of $f$ on $\timedep{\sigma}$ are bounded by the minimum and maximum coefficients of $f$ when expressed in the Bézier basis.

To obtain the vector of Bézier coefficients $f^\bezpoly$ of $f$, we first compute its vector of Lagrange coefficients $f^\lagpoly$, which can be obtained by simply evaluating $f$ at the domain points; then we premultiply $f^\lagpoly$ with a change of basis matrix $\transition{\lagpoly}{\bezpoly}$ that we shall call \emph{transformation matrix}, which is described in detail in \cref{app:bases}.

Let $\timedep{\indexset}_{\sigma}^m$ be the set of indices of the control points of $f$ and $\timedep{\cornerset}_{\sigma}^m\subset\timedep{\indexset}_{\sigma}^m$ 
be the set of indices corresponding to the corners of $\timedep{\sigma}$ at time $1$.
Since the Bézier basis is interpolating at the corners of the domain (i.e. $\beta_j = f(\gamma_j)$ for all $j\in\timedep{\cornerset}_{\sigma}^m$), we define the minimum inclusion function as
\begin{equation}
\label{eq:minclusion-bezier}
\minclusion f(\timedep{\sigma}) = [\min_{i\in\timedep{\indexset}_{\sigma}^m}\beta_i,\min_{j\in\timedep{\cornerset}_{\sigma}^m}\beta_j].
\end{equation}
Therefore, if all entries of $f^\bezpoly$ are positive we know the element is valid everywhere, and if any of the corner entries is non-positive we know that the element is invalid at the end time. 
Otherwise, the interval returned by the inclusion function contains zero, and we need to refine the search by subdividing $\timedep{\sigma}$. 

The subdivision of $\timedep{\sigma}$ is performed via another set of change of basis matrices, dubbed \emph{subdivision matrices}, $\transition{\bezpoly}{\bezpoly}^q$, for $q=1,\dots, Q$. Premultiplication of $f^\bezpoly$ by these matrices gives a Bézier representation of $f$ on a smaller portion of the domain, which can be used to compute tighter bounds local to each subdomain. These matrices are defined in \cref{app:bases}.

\paragraph{Time-only refinement}
Bisection in the time dimension only is performed analogously by multiplication of $f^\bezpoly$ with two \emph{time subdivision matrices} $\transition{\bezpoly}{\bezpoly}^-$ and $\transition{\bezpoly}{\bezpoly}^+$, also described in \cref{app:bases}.

\subsection{Robust Computation}
\paragraph{Rational Precomputation of matrices}
All the transformation and subdivision matrices $\transition{\lagpoly}{\bezpoly}$, $\transition{\bezpoly}{\bezpoly}^q$, and $\transition{\bezpoly}{\bezpoly}^\pm$
are only dependent on element type (tetrahedron, hexahedron, etc.) and order, and as such can be precomputed offline. Since we want to minimize the accumulation of error in our computations, and all entries of these matrices are rational, we construct these matrices using exact rational arithmetic for each element type and order. The outcome is a rational matrix, which we convert to intervals by rounding the two endpoints outward if the exact value cannot be represented as a floating point number. The resulting interval is guaranteed to contain the exact value of the fraction while being as tight as possible.

\paragraph{Intervals}
The input to our check is the set of Lagrange control points for the elements of a mesh, represented in floating point. Each floating point coordinate is converted to a singleton interval (i.e. an interval with zero width) and all subsequent operations are performed in interval arithmetic with conservative rounding.

\paragraph{Summary}
The combination of rational precomputation and interval arithmetic ensures that our algorithm is conservative while maintaining a low computational cost (\cref{sec:results}): the rational precomputation is performed only once offline and does not affect runtime, while the use of interval arithmetic adds a minor ($\sim\!\!2\times$) overhead over a direct floating point implementation.

\subsection{Acceleration}

\paragraph{Global Queries}
In practical applications, one is often interested in the maximum time for which \emph{all} elements are valid. 
We refer to this as a \emph{global} dynamic query and give a strategy to accelerate it.

After an invalid element has been found (with estimated valid time step $\intlo{t^*}$), it becomes unnecessary to validate the other elements at later times: $\intlo{t^*}$ will anyhow be the maximal allowed step. 
It might be tempting to terminate early as soon as an inverted element has been found. However, this is not conservative as some other elements may still have a lower $\intlo{t^*}$.
We thus keep track of the smallest value of $\intlo{t^*}$ found in previous checks and leverage the fact that $\intlo{t^*}$ can never decrease in \cref{alg:bezierDynamic} to stop computation on an element as soon as its estimate for $t^*$ exceeds the running minimum.

The order in which elements are processed matters for global queries: it is beneficial to process elements that are most likely invalid first, as it will provide higher opportunities for this pruning strategy to be effective. For this reason, we first sort the polynomials according to their constant term, in ascending order.
For the order 3 Armadillo mesh in \cref{fig:dataset}, this strategy improves the total running time by about 60\% over running the queries individually; whereas the speedup is less relevant for the order 2 mesh (about 12\%). 

\paragraph{Parallelization}
Validity checks for meshes are trivially parallelized by processing elements in batches. To avoid any synchronization between different threads, every batch of queries assigned to a thread $i$ does an independent sorting and keeps its own running minimum $\intlo{t^*}_i$ to use as an early termination condition, as explained in the previous paragraph.

\paragraph{Precomputation of Jacobian determinant}
The input to the subdivision procedure is a Lagrange representation of the Jacobian determinant polynomial of the element. This only depends on the shape and order of the element, as well as its control points.

For each element type and order combination, we symbolically compute the expression of each Lagrange coefficient in terms of the control point coordinates, and remove common subexpressions with CSE \cite{Muchnick1997} (we use the implementation in 
SymPy \cite{sympy}). This approach increases the compilation time but provides dramatic runtime performance boosts: for the order 3 Armadillo dataset, we get a speedup of about $20\times$.

\section{Application to simulation}
\label{sec:applications}

Incremental potential time-stepping \cite{Kane2000} is becoming popular in graphics \cite{Li2020IPC} and biomechanics \cite{Martin2024} due to its robustness to extreme deformation and contact\cite{CIPC2021,MedialIPC2021,IDP2021,RigidIPC2021,ABD2022,PDIPC2022,BFEMP2022,Chen2022,highorderIPC,Li2023GPUClothIPC,Lan2023,DiffIPC2024,GIPC2024,PIPC2024,Duo2024,Fang2024}.
We briefly summarize the approach here, without contact handling, as it is relevant to motivate the need for a continuous dynamic positivity check in physical simulation: as part of this overview, we will show that the check alone is insufficient, as IPC also requires a \emph{consistent} invalidity-aware quadrature rules, which we introduce in Section \ref{sec:custom-quadrature}.

\subsection{Continuous Validity in Simulation}
\label{sec:custom-quadrature}
\label{sec:continuous}

The updated displacement $u^{t+1}$ of an object at the next time step is computed solving an \emph{unconstrained} non-linear energy minimization:
\begin{equation}
  u^{t+1} = \argmin_u \> E(u, u^t, v^t),
  \label{eq:ipc-in-ho}
 \end{equation}
where %
$u^t$ is the displacement at the step $t$, $v^t$ is velocity, and $E(u, u^t, v^t)$ is a time-stepping Incremental Potential \cite{Kane2000}.
We refer to \cite{Li2020IPC} for more details. %

For common non-linear material models, this potential is infinite when an element has a negative Jacobian, as the Jacobian determinant appears in the denominator of the expression. A physically valid trajectory cannot reach a state with infinite potential: however, this is a challenging condition to enforce in practice.

\begin{figure}
    \centering
    \includegraphics[width=0.7\linewidth]{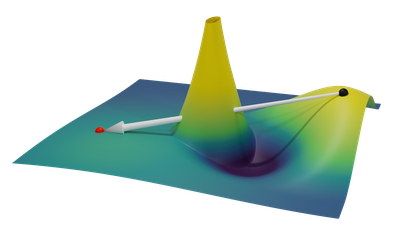}
    \caption{Potential landscape and one descent step from the black point. The descent step to the red point is invalid since it crosses an invalid region with infinite potential.}
    \label{fig:landscape}
\end{figure}

\paragraph{Line Search} The potential $E$ is minimized with a descent algorithm (gradient descent or Newton), %
which computes a local approximation of a descent direction: this approximation might, for a finite step length, cross a region with infinite potential (\cref{fig:landscape}). This is a typical challenge in collision detection \cite{wang_large-scale_2021}, but is rarely considered for the elastic potential -- the only work we know that considers this problem, in 2D only, is \cite{Smith:2015}. 
This challenge can be solved using a continuous inversion check within the line search, which is the focus of our work. To the best of our knowledge, state-of-the-art IPC solvers \cite{Li2020IPC,polyfem} use a static check instead of a continuous one, which cannot guarantee trajectory validity.

\begin{figure}
    \centering
    \includegraphics[width=.45\linewidth]{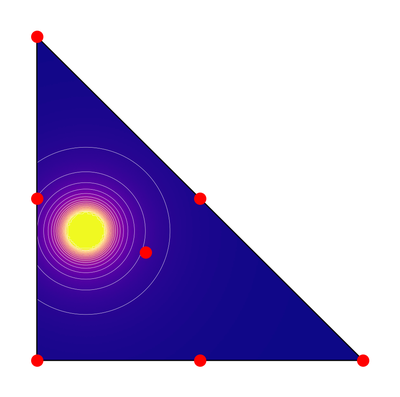}
    \includegraphics[width=.45\linewidth]{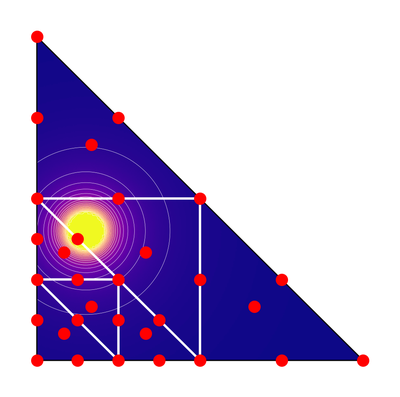}
    \caption{The function is infinite in the bright yellow region, and its integral over the whole triangle is also infinite. Numerical integration using a fixed quadrature rule (left) erroneously produces a finite value. Our adaptive quadrature technique (right) puts quadrature points in the infinite-valued region and correctly captures the behavior of the function.}
    \label{fig:quadrature}
\end{figure}

\paragraph{Quadrature}
Non-linear elastic potentials cannot be integrated exactly with numerical quadrature (as they are not polynomials), leading to unbounded errors for diverging potentials. 
We show in \cref{fig:quadrature} an example of an element with an infinite potential integrated with both a standard fixed quadrature rule and the adaptive quadrature derived by our algorithm: only in the second case does the numerical integration correctly diverge. 
The use of a fixed quadrature leads to solver failures %
as the direction computed using quadrature is not a descent direction and might thus block the progress of the solver.

\subsection{Invalidity-Aware Quadrature Rules}
\label{sub:our-quadrature}
Our algorithm can output additional information to generate adaptive quadrature schemes: The goal is to produce a set of quadrature points for the static element $\sigma$ such that at least one of the points would end up in an invalid region at time $t^*$, correctly making the integral computed using quadrature diverge at that time.

\paragraph{Tracking Subdivisions}
In \cref{alg:bezierDynamic}, each subdomain keeps track of the sequence of subdivisions of $\timedep{\sigma}$ that were taken to produce it. However, since we only require a quadrature rule for $\sigma$ and not $\timedep{\sigma}$, it is sufficient to keep track of spatial subdivisions: this means that when an element is subdivided in time only using $\psi^+$ and $\psi^-$, the sequence of the children will match that of the parent; and when an element is subdivided in all dimensions using the subdivision scheme $\psi^q$ for $\timedep{\sigma}$, pairs of subdomains that span the same region of space at different times will share the same sequence.

To produce the adaptive quadrature rule, our algorithm uses the sequence of spatial subdivisions of the element whose minimum inclusion function reduced $\inthi{t^*}$ for the last time, or equivalently, the element that found the earliest invalidity.
Unless the early exit condition for maximum depth is triggered, such invalidity is guaranteed to be in $[\intlo{t^*}, \intlo{t^*}+\delta]$.
In the very rare cases when the algorithm fails to find an invalid point and gives up earlier, we instead return the subdivision sequence of the element with the deepest hierarchy, which is likely to be very close to an invalid region. In this case, the optimization step will still be guaranteed to be valid for its duration.

\paragraph{Adaptive Quadrature}
This information is then used to partition the static element by recursively subdividing it using the very same sequence of subdivisions, as in \cref{fig:quadrature} (right). 
A standard quadrature rule is applied to every subelement of $\sigma$, and the integral is evaluated as the sum of integrals on all subdomains.
However, in order to guarantee that the newly placed quadrature points would indeed intersect an invalid region in the full time step, it is required that the quadrature rule contain all points in the sampling set $\sampleset{D}$ used to compute the minimum inclusion function, projected onto $\sigma$; for example, if $\sampleset{D}$ is the set of corners of $\timedep{\sigma}$, the selected scheme must place quadrature points at the corners of $\sigma$.

\section{Results}
\label{sec:results}

Our algorithm is implemented in C++, using 
PolyFEM \cite{polyfem}
for finite element (FE) system construction, IPC Toolkit \cite{IPCToolkit} for evaluating IPC potentials and collision detection,
Pardiso \cite{Alappat2020Recursive,Bollhofer2020State,Bollhofer2019Large} for
the large linear systems in our global Newton solves, \cite{ipred_lib,attene20} for interval computation, GMP \cite{Granlund12} for rational computation, and OpenMP \cite{openmp} for parallelization.

The simulation experiments are run on a cluster node with an Intel Cascade Lake Platinum 8268 processor limited to 16 threads and 32Gb of memory.
The benchmark experiments are run single-threaded on a laptop computer with an AMD Ryzen 7 4000 processor and 16GB of memory.

Our reference implementation, used to generate all results, and our evaluation datasets will be released as an open-source project. 
 
\begin{table*}
	\begin{tabular}{l|l|cc|c|r|r|r|r|r|r|r|r|r}
		Dataset & Element & $n$ & $s$ & $p$ & \multicolumn{3}{c|}{Element count} & \multicolumn{6}{c}{Time per element ($\mu$s)}\\
            & & & & & tot & val & inv & avg & avg val & avg inv & med & max & std  \\
		\hline
            Kangaroo & Triangles & 2 & 2 & 1 & 172800 & 146248 & 26552 & 1.60 & 1.32 & 3.10 & 1.40 & 205.40 & 0.85 \\
                     && & & 2 & 172800 & 145602 & 27198 & 2.14 & 1.69 & 4.54 & 1.89 & 42.53 & 1.10 \\
                     && & & 3 & 172800 & 145623 & 27177 & 3.83 & 3.10 & 7.75 & 3.28 & 67.26 & 1.86 \\
                     && & & 4 & 172800 & 145613 & 27187 & 9.28 & 7.94 & 16.45 & 7.96 & 309.19 & 4.03 \\
            Bar 2D & Quadrangles & 2 & 1 & 1 & 112887 & 14816 & 98071 & 3.93 & 2.95 & 4.07 & 3.84 & 10203.74 & 34.23 \\
                     && & & 2 & 83653 & 75085 & 8568 & 74.67 & 7.89 & 659.96 & 2.93 & 2046480.70 & 10296.53 \\
            Armadillo & Tetrahedra & 3 & 3 & 1 & 54985 & 51605 & 3380 & 1.56 & 1.44 & 3.43 & 1.47 & 21.58 & 0.53 \\
                     && & & 2 & 54985 & 48599 & 6386 & 9.90 & 7.34 & 29.41 & 5.31 & 6726.90 & 40.37 \\
                     && & & 3 & 54985 & 47988 & 6997 & 362.79 & 130.92 & 1953.05 & 44.00 & 78411.99 & 1593.81 \\
            Bunny & Tetrahedra & 3 & 3 & 1 & 19800 & 19561 & 239 & 1.47 & 1.45 & 3.51 & 1.40 & 49.38 & 0.63 \\
                    && & & 2 & 19800 & 19058 & 742 & 7.27 & 5.91 & 42.22 & 5.31 & 1729.90 & 24.56 \\
                     && & & 3 & 19800 & 18954 & 846 & 175.41 & 65.97 & 2627.34 & 43.02 & 60907.39 & 1294.73 \\
            Bar 3D & Hexahedra & 3 & 1 & 1 & 56031 & 24918 & 31113* & 325.33 & 11.10 & 576.99 & 16.55 & 6650064.32 & 39487.14 \\ 
                    & Prisms & 3 & 2 & 1 & 82403 & 56913 & 25490* & 246.64 & 8.22 & 778.96 & 2.38 & 2407415.47 & 16401.57 \\ 
	\end{tabular}
	\caption{Results of our continuous validity test on 2D and 3D datasets element types and orders. We report the number of elements (total, valid on the whole interval, invalid at some time) and processing time in microseconds per element (average, average for valid elements, average for invalid elements, median, maximum, standard deviation).
      On some datasets, marked with an asterisk (*), the algorithm ``gives up'' on very few elements and returns a conservative answer: see details in \cref{sec:limitations}. In all other cases, the algorithm reaches target precision on all elements. 
      }
	\label{tab:exp}
\end{table*}

\subsection{Benchmark}

\begin{figure}
    \centering
    \includegraphics[width=\linewidth]{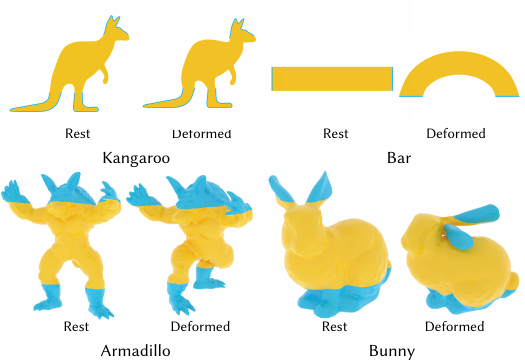}
    \caption{Initial and final frames of simulations from which queries are exported. The parts in cyan are used as handles to twist and compress the model while the deformation occurs in the parts in yellow.}
    \label{fig:dataset}
\end{figure}

\paragraph{Collection and Ground Truth} We collect queries of element inversion checks from elastodynamic simulation data using IPC~\cite{Li2020IPC} and the Neo-Hookean elasticity model. We pick two 2D models and two 3D models shown in \cref{fig:dataset}. We bend the bar and compress the kangaroo in 2D, whereas for 3D models we twist them by $90^\circ$ while compressing them by $20\%$. %

\begin{figure}
    \centering
    \includegraphics[height=0.44\columnwidth]{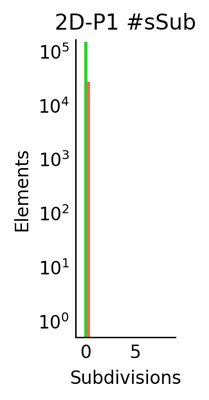}
    \includegraphics[height=0.44\columnwidth]{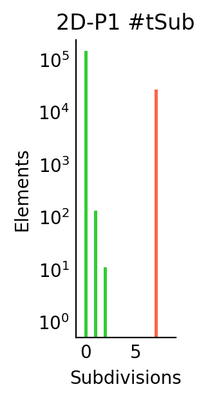}
    \includegraphics[height=0.44\columnwidth]{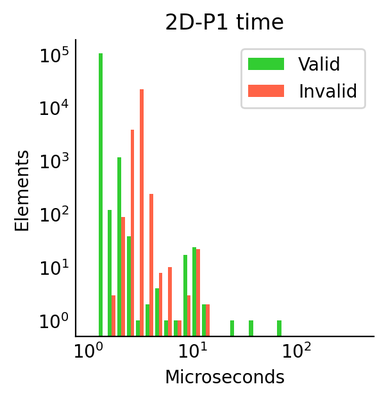}\\
    \includegraphics[height=0.44\columnwidth]{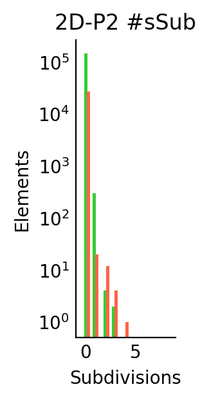}
    \includegraphics[height=0.44\columnwidth]{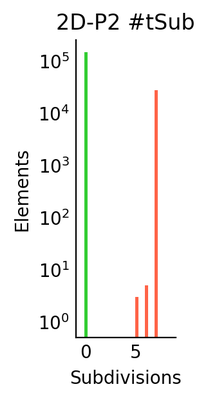}
    \includegraphics[height=0.44\columnwidth]{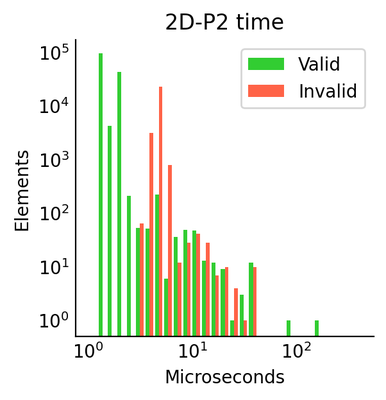}\\
    \includegraphics[height=0.44\columnwidth]{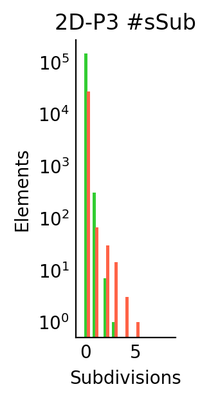}
    \includegraphics[height=0.44\columnwidth]{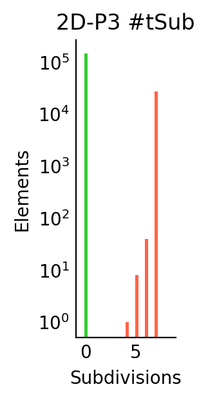}
    \includegraphics[height=0.44\columnwidth]{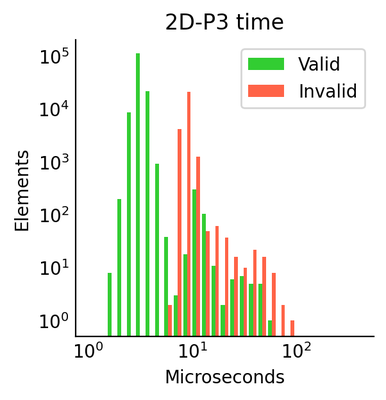}\\
    \includegraphics[height=0.44\columnwidth]{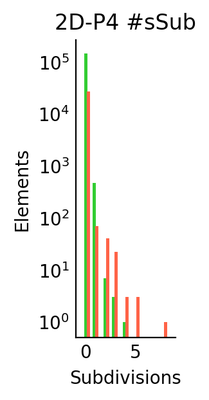}
    \includegraphics[height=0.44\columnwidth]{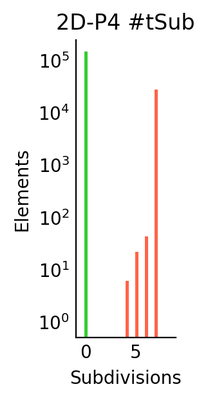}
    \includegraphics[height=0.44\columnwidth]{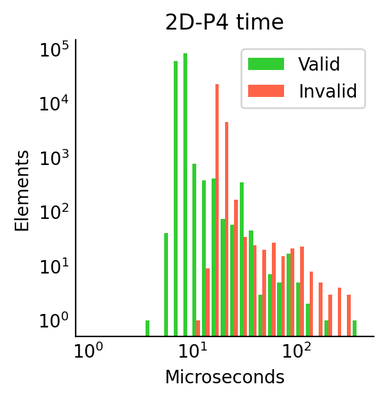}
    \caption{Statistics for the Kangaroo datasets. Top to bottom: elements of order 1, 2, 3, 4; Left to right: number of space subdivisions ($\#$ssub), time subdivisions ($\#$tsub), and time to test an element. Green valid elements; red invalid ($t^*<1$) elements.}
    \label{fig:statistics-2D}
\end{figure}

\begin{figure}
    \centering
    \includegraphics[height=0.44\columnwidth]{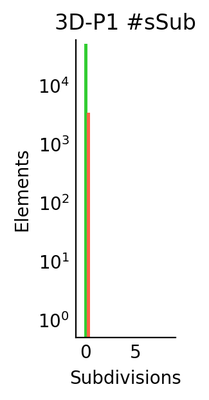}
    \includegraphics[height=0.44\columnwidth]{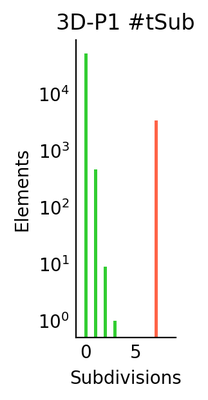}
    \includegraphics[height=0.44\columnwidth]{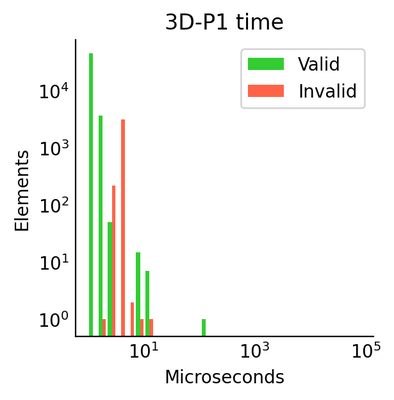}\\
    \includegraphics[height=0.44\columnwidth]{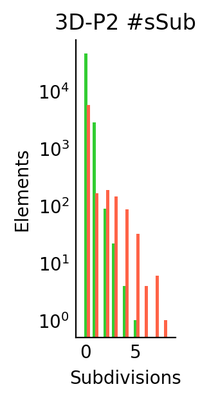}
    \includegraphics[height=0.44\columnwidth]{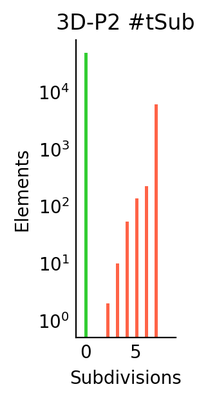}
    \includegraphics[height=0.44\columnwidth]{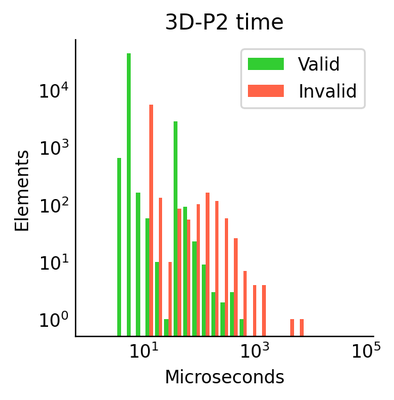}\\
    \includegraphics[height=0.44\columnwidth]{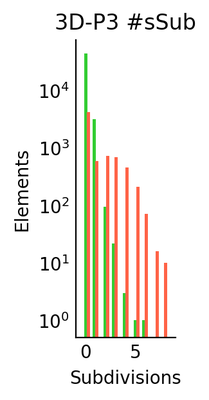}
    \includegraphics[height=0.44\columnwidth]{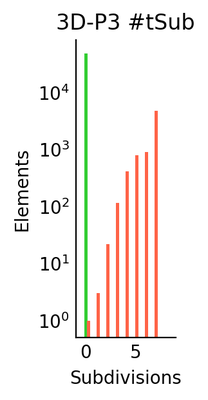}
    \includegraphics[height=0.44\columnwidth]{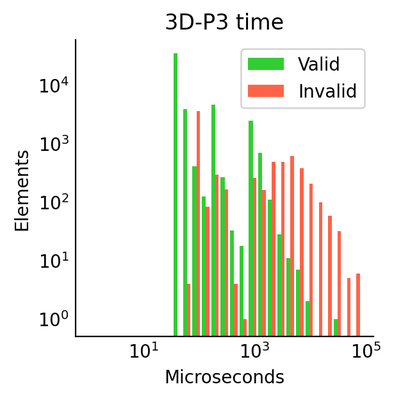}
    \caption{Statistics for the Armadillo datasets. Top to bottom: elements of order 1, 2, 3; Left to right: number of space subdivisions  ($\#$ssub), time subdivisions  ($\#$tsub), and time to test an element. Green valid elements; red invalid ($t^*<1$) elements.}
    \label{fig:statistic-3D}
\end{figure}

Our benchmark contains 172000 2D queries from the Kangaroo model (orders 1 through 4), 54985 3D queries from the Armadillo model (orders 1 through 3), 19800 3D queries from the Bunny model (orders 1 through 3), 130291 queries for the Bar model of order 1, and 31879 queries for the Bar model of order 2 (\cref{tab:exp}). 

\paragraph{Correctness}
To validate correctness we compute the times of inversion by symbolically computing the roots of $|J_f|$ with minimum $t$ using Mathematica \cite{Mathematica} (see \cref{app:mathematica}). We restrict the ground truth to order 1 and 2 for triangles, and order 1 for quads and tetrahedra, due to the limitations of symbolic solvers: for a higher-order basis, Mathematica could not return a result within 6 hours on some of the elements. To the best of our knowledge, ours is the first dataset containing conservative times of inversion. Our algorithm correctly detects all invalid elements and returns conservative answers for the inversion times.

\paragraph{Efficiency}
The average per-element cost of our algorithm increases with the degree, from around 1.5$\mu$s for order 1 (both in 2D and 3D) to around 10$\mu$s for order 4 in 2D and 300$\mu$s for order 3 in 3D.
Invalid elements are more expensive to process on average, as they require subdivisions until precision is reached, whereas most valid elements can be resolved in a single iteration if all their Bézier coefficients are positive. See \cref{tab:exp} and \cref{fig:statistics-2D,fig:statistic-3D} for details.

When restricted to the static case (i.e., check the validity of an element at a given time), our algorithm becomes considerably faster than the dynamic one (by a factor of about $50\times$ on the order 3 Armadillo), and it has a running time slightly faster ($2\times$) than the non-conservative static baseline used in PolyFEM and FEBio, which consists of checking the sign of $|J_x|$ obtained with (inexact) floating point computations only at quadrature points thanks to our optimizations.
See \cref{tab:static} for details.

\subsection{Comparisons}
\label{sub:comparisons}

We are not aware of any other algorithm that provides a continuous validity check for elements of arbitrary order, so we compare it with algorithms that solve only a subset of the problem.

\paragraph{Linear Continuous}

For the special case of triangular elements with linear basis, \cite{Smith:2015} proposes to use a symbolic solver to find the roots of $|J_f|$. While extremely efficient (1$\mu$s on average), this approach can fail to produce correct results due to numerical errors: on the 26552 invalid elements in the linear Kangaroo dataset (\cref{fig:dataset}), their method fails to detect inversions 13324 times, producing a  time step larger than the ground truth maximum. Attempting to be conservative with a "large" numerical threshold of $10^{-5}$ still fails in 1563 tests ($\sim\!\!5\%$).

This method is limited to linear triangles, and cannot be extended to other elements or degrees due to its reliance on closed-form expressions for the roots. %

\paragraph{Static High Order}

For the special case of static validity check for elements of arbitrary type and order, \cite{johnen_geometrical_2014} introduces a method based on adaptive subdivision. We implemented and tested their method in double floating-point precision, and while more efficient than our static approach (on the 3D Armadillo model of order 3 our method takes $3\mu s$ longer per element on average, see \cref{tab:static}), it is not conservative. We provide one example of an inverted element incorrectly detected as valid by our implementation of the algorithm in \cite{johnen_geometrical_2014} in \cref{app:gmsh}, whose code is attached to the submission. 

\begin{table}
	\begin{tabular}{l|r|r|r|r|r}
		Static algorithm & \#val & \#inv & \#und & avg $\mu$s & med $\mu$s \\
		\hline
            Quadrature Points & 48214 & 6771 & - & 16 & 16 \\
            Interval Bisection & 46281 & 6561 & 2143 & 1400 & 23 \\
            FP Bézier (no optim.) & 48050 & 6935 & 0 & 71 & 78 \\
            Ours (no optim.) & 48050 & 6935 & 0 & 86 & 95 \\
            FP Bézier & 48050 & 6935 & 0 & 5 & 5 \\
            \textbf{Ours} & 48050 & 6935 & 0 & 8 & 8 \\
	\end{tabular}
    \caption{Comparison of methods for static validity checks, performed at $t=1$ on the order 3 Armadillo dataset.
    We list the number of detected valid and invalid elements, the number of elements for which the test was undecided, and the average and median computation times per element. 
    Sampling at quadrature points is fast, but incorrectly classifies several invalid elements as valid; bisection with a robust interval-based inclusion function is fast on "easy" elements, but struggles a lot with nearly-inverted elements and fails to classify several of them; our implementation of the Bézier refinement based inclusion check from \cite{johnen_geometrical_2014} and our conservative method give the same results; our method is guaranteed correct at a slight performance cost.
    Our precomputations (last two rows) decrease the computation time by at least an order of magnitude.}
	\label{tab:static}
\end{table}

\begin{figure}
    \centering
    \includegraphics[width=\linewidth]{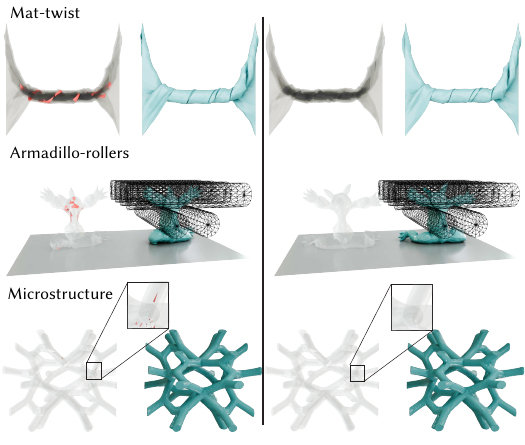}
    \caption{Simulation examples in \cite{highorderIPC} with (right) and without (left) our inversion check. The high-order elements with non-positive Jacobian points are shown in red. On the left, the numbers of flipped elements are 20, 45, and 24 from top to bottom. Our method guarantees the positivity of Jacobian.}
    \label{fig:simulation-flip}
\end{figure}

\begin{table*}[htb]
    \centering
    \begin{tabular}{c|c|c|c|c|c|c|c|c|c|c}
         &  & \multicolumn{2}{c|}{Order} & & & Mem & \multicolumn{4}{c}{Timing (\unit{\sec})} \\
        & dim & geom & soln & \# Cells & \# Steps & (\unit{GB}) & time step & check per time step  & check avg & baseline \\ \hline
        Beam-twist & 3 & 1 & 2 & 1740 & 400 & 1.4 & 16 & 5.7 & 0.067 & 7.6 \\
        Ring-twist & 2 & 1 & 2 & 1136 & 100 & 0.7 & 0.72 & 0.16 & 0.011 & 0.15 \\
        Mat-twist & 3 & 1 & 2 & 2166 & 250 & 1.3 & 10 & 2.9 & 0.069 & 6.9 \\
        Armadillo-rollers & 3 & 1 & 2 & 5978 & 230 & 3.3 & 114 & 27 & 0.19 & 111 \\
        Microstructure & 3 & 4 & 2 & 6414 & 25 & 2.9 & 531 & 472 & 5.6 & 52 \\
    \end{tabular}
    \caption{Simulation Statistics. Columns from left to right: simulation dimension, geometric and solution bases orders, number of cells, number of time steps, peak memory usage, the average time of the simulation per time step, the average time of our check per time step, the average time of each check over the entire mesh, and average time of the simulation per time step using the quadrature point check in PolyFEM instead of ours.}
    \label{tab:sim_stat}
\end{table*}

\begin{figure}
    \centering
    \includegraphics{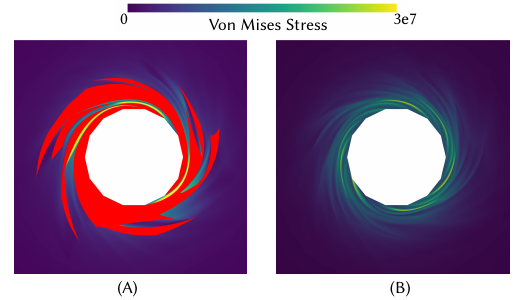}
    \caption{Von Mises stress distribution of Ring-twist in \cref{fig:teaser}. The NaN is shown in red. (A) Without our method, NaN appears on elements with flipped points. (B) With our Jacobian check and adaptive quadrature, the stress is everywhere finite. }
    \label{fig:nan-stress}
\end{figure}

\subsection{Elastodynamic Simulation}
\label{sec:results:elasto}

We integrate our check within the PolyFEM software \cite{polyfem} and use it to reproduce a selection of the bundled elastodynamic simulations. We report our findings as we integrated the check, since they highlight some fundamental issues with existing high-order FE solvers and the non-linear material models commonly used in graphics and engineering. 

\paragraph{Potential Formulation}

In the simulation, we minimize the Neo-Hookean energy with Newton's method. The Neo-Hookean energy density has the form of 
\begin{equation}\label{eq:neohookean}
    w_e(F) := \frac{\mu}{2}  (\text{Tr}[F F^T] - 2 - 2\log (\det F)) + \frac{\lambda}{2} \log^2 (\det F), 
\end{equation}
where $F$ is the deformation gradient matrix ($2\times 2$ or $3\times 3$), $\lambda$ and $\mu$ are Lam\'e parameters. Suppose the Jacobian of the deformed element and the rest element is $J_d$ and $J_r$, the deformation gradient is given by
$$
F=J_d (J_r)^{-1}.
$$
Suppose the rest element is valid, i.e., $\det J_r > 0$; then the positivity of $\det F=\det J_d / \det J_r$ depends solely on $\det J_d$, i.e. the Jacobian determinant that we are checking. As long as $\det J_d>0$, \cref{eq:neohookean} is valid.

\paragraph{Baseline Simulator}
Our baseline simulator is PolyFEM \cite{polyfem}, using the convergent IPC formulation \cite{li2023convergent} on tetrahedral meshes, quadratic Lagrangian bases and the Neo-Hookean material model \cite{ogden2013non}. PolyFEM uses only a static inversion check on the quadrature points: in \cref{fig:teaser,fig:simulation-flip} we show that the final result contains many invalid elements leading to NaN in stress (\cref{fig:nan-stress}A). These invalid elements are not detected using the check on quadrature points alone, but are correctly identified by our conservative check (\cref{fig:nan-stress}B). %

\paragraph{Conservative Line-Search Only}
Replacing the static sampling invalidity check with our conservative and continuous check without adaptive quadrature leads to convergence issues in the solver. When the elements become close to inversion, the negative gradient direction is not a descending direction. This happens due to the error in the numerical integration of the potential, that can be solved by using our adaptive quadrature method. All the simulations in our experiments suffer from this issue: the simulation halts after several steps (twist-beam fails at time step 145, armadillo-rollers at 33, mat-twist at 141, ring-twist at 28, and the microstructure at 7).

\paragraph{Conservative Line-Search + Adaptive Quadrature}
This solution works but dramatically increases the number of non-linear iterations needed in the Newton method. The remaining issue is that, while the Neo-Hookean energy density is infinite at the point where the Jacobian determinant is zero, it grows at a slow rate. 
Mathematically, when the Jacobian determinant is exactly zero at one point and positive everywhere else, the integral of the Neo-Hookean energy still may be finite: consider the integral of a 1D function $f(x)=-\log(|x|)$ on the interval $[-1,1]$, although it has a singularity at $x=0$, the integral is
$$
\int_{-1}^1 -\log(|x|)dx=-2\int_0^1 \log(x)dx=2<\infty
$$
This is undesired since it leads to NaN in \cref{eq:neohookean} and stress evaluation, causing failure of the simulation. This problem can be mitigated by adding an additional barrier term $\frac{\mu}{\det F}$ to \cref{eq:neohookean}. While this solution formally fixes the problem (and also practically fixes it in our experiments), it makes the potential harder to minimize while changing the material model: we are not aware of any analysis of this problem, and we believe it is an exciting avenue for future work.

\paragraph{Observations.}
With these modifications, PolyFEM produces results without invalid elements. 
We note that given that the material models are infinite for invalid elements, the existing non-conservative approaches often create solutions that are non-physical. A more detailed study of the effect of these errors could be an interesting venue for future work.

A second observation is that the issue with the slow growth of the Neo-Hookean potential (and others, such as Mooney-Rivlin) is likely due to the misuse of these models for deformations that are outside of the regime they are designed to handle. It would be interesting to carefully study experimentally how accurate these material models are under extreme deformations, and see if it is possible to design other material models which are numerically more suitable for interior point optimization.

\paragraph{High-Order IPC}
We reproduce simulation examples that use high-order finite elements in \cite{highorderIPC}, including Mat-twist, Armadillo-rollers, and Microstructure (\cref{fig:simulation-flip}). 
In \cref{fig:teaser}, we include two more examples: Beam-twist and Ring-twist. In the Beam-twist, we apply Dirichlet boundary conditions on the two sides of the beam, rotate one side, and keep the other side fixed; in the Ring-twist, we apply Dirichlet boundary conditions to rotate the inner circle of the ring with constant speed and allow the outer circle free to move. In \cref{fig:simulation-flip}, the simulation results in \cite{highorderIPC} have flipped elements since the simulator only checks Jacobian at quadrature points in each element, while with our conservative check and adaptive quadrature, there is no flipped element. We report the statistics in \cref{tab:sim_stat}. For quadratic elements, the runtime of our method is at worst comparable to the solve time and can be as fast as 23\% of the solve time; for quartic geometric elements and quadratic solutions, our method is much slower than the solve time, since the solve is on quadratic elements while the Jacobian check should be performed on the quartic elements. %

\section{Conclusions}\label{sec:concluding}

We introduced a formulation for continuous inversion test and a corresponding conservative and efficient algorithm. Our solution addresses an open problem in existing finite element solver and parametrization algorithms, increasing robustness and providing, for the first time, a guarantee for interior point solvers to stay within the space of valid elements. While the issue does appear for linear elements, invalid elements are more commonly present in existing algorithms that use high-order bases. 
We believe our algorithm, and its reference implementation, will be a drop-in replacement for existing non-conservative checks used in many graphics algorithms that will increase robustness for a minor performance cost.

\paragraph{Limitations} \label{sec:limitations}

The cost of our algorithm increases with the polynomial degree, especially for tensor product elements. 
While this is compatible with many applications in graphics and engineering (up to cubic tetrahedra, linear hexahedra/prisms), it is not scalable to very high-order FEM, where elements of order 20 or more are routinely used.

Additionally, similarly to all bisection algorithms, our approach might require a lot of refinement in certain queries where it is challenging to find the roots: while we could not find such a case for simplicial elements, our dataset contains some tensor product elements (57 hexahedra and 26 prisms) for which the algorithm exceeds the available memory on the personal machine where the benchmark was run, and must be stopped early via the $l_{\max}$ parameter (set to a maximum subdivision depth of 7): our algorithm still returns a conservative answer.
Altering the subdivision strategy mitigates this issue and allows the algorithm to converge on the whole dataset: exploring better heuristics for subdivision may be a fruitful endeavor for improvement.

\paragraph{Future Work}

Thanks to our approach, we discovered a previously unreported numerical problem with existing material models under extreme compression, which was obfuscated by inaccurate quadrature rules. 
We believe evaluating these material models in extreme compression regimes would be an interesting avenue for future work. Additionally, we discovered that high-order elastodynamic FE solvers often introduce inverted elements in their solutions: their effect on simulation accuracy requires further investigation, which is now possible since our test can detect them.

To further reduce the computational cost, it would be interesting to investigate the possibility of parallelizing individual queries or adapting our approach for massively parallel graphic processing units.

\bibliographystyle{ACM-Reference-Format}
\bibliography{biblio.bib}

\appendix
\section{Langrange and Bézier forms of $|J_x|$}
\label{app:bases}

Let $f:\sigma_s^n\longrightarrow\reals$ be a polynomial of order $p$ defined on a standard element according to \cref{def:element}. 
We detail the representation of $f$ in Lagrange and Bézier forms and the conversions between these two representations. 

\paragraph{Lagrange form}
Let $\domainpts_{\sigma}^p = (1/p)\ints^n \cap \sigma$ be a grid of uniformly distributed domain points of $\sigma$, and $\indexset_\sigma^p$ be its set of indices (\cref{fig:domains}). 
The Lagrange basis of order $p$ consists of $|\domainpts_{\sigma}^p|$ order $p$ polynomials such that for each point $\gamma_i\in\domainpts_{\sigma}^p$, $\lagpoly^p_{i}(\gamma_j)=\delta_{ij}$.
A function $f$ is represented in the Lagrange basis as:
\begin{equation}
\label{eq:lagrange-map}
	f(\xi) = \sum_{i\in\indexset_\sigma^p} y_i \lagpoly^p_{i}(\xi),
\end{equation}
where $y_i = f(\gamma_i)$ for $i\in\indexset_\sigma^p$.

\paragraph{Bézier form}
The same function $f$ can be expressed equivalently in Bézier form by using a Bernstein basis on the same set $\domainpts_{\sigma}^p$ of domain points:
\begin{equation}
\label{eq:bezier-map}
	f(\xi) = \sum_{i\in\indexset_\sigma^p} \beta_i \bezpoly^p_{i}(\xi),
\end{equation}
where the $\bezpoly^p_{i}$ are Bernstein polynomials of order $p$ and the $\beta_i$ are their corresponding control coefficients. 
Unlike the Lagrange case, $\beta_i$ equals $f(\gamma_i)$ only at the corners of $\sigma$.
However, the graph of $f$ is contained in the convex hull of points $(\gamma_i,\beta_i)$, providing a simple way to bound $f$ from below and above at all points of $\sigma$ \cite{Farin:2001}.

\paragraph{Transformation matrix}
Let us denote $f^\lagpoly$ the vector consisting of all the $y_i$, and likewise, $f^\bezpoly$ the vector consisting of all the $\beta_i$, for $i\in\indexset_\sigma^p$. 
We can convert between the two representations through transformation matrices \cite{johnen_geometrical_2014}:
\begin{equation}
\label{eq:conversion}
f^\lagpoly=\transition{\bezpoly}{\lagpoly}f^\bezpoly \qquad f^\bezpoly=\transition{\lagpoly}{\bezpoly}f^\lagpoly.
\end{equation}
Such matrices depend only on the reference element $\sigma_s^n$ and the order $p$ thus can be computed once for each element type and order.
Matrix $\transition{\bezpoly}{\lagpoly}$ is easily computed by evaluation of Bernstein polynomials on $\Gamma_{\sigma}^p$:
\begin{equation}
	(\transition{\bezpoly}{\lagpoly})_{ij} = \bezpoly^p_j(\gamma_i) \quad \forall i, j \in \indexset_\sigma^p
\end{equation}
and $\transition{\lagpoly}{\bezpoly}$ is the inverse of $\transition{\bezpoly}{\lagpoly}$.

\paragraph{Subdivision matrices}
For each $q$, we first build a transformation matrix from the Bézier basis to the Lagrange basis that interpolates the domain points $\psi^q(\Gamma(m))$ of the $q$-th subdomain. 
This is analogous to building matrix $\transition{\bezpoly}{\lagpoly}$, by sampling the Bézier basis on $\psi^q(\Gamma(m))$ instead of $\Gamma(m)$:
\begin{equation}
\label{eq:q-BtoL}
	(\transition{\bezpoly}{\lagpoly}^q)_{ij} = \bezpoly^m_j(\psi^q(\gamma_i)) \quad \forall i, j \in \indexset_\sigma(m).
\end{equation}
Then we multiply it with the Lagrange-to-Bézier matrix to build
\begin{equation}
	\label{eq:subdiv-matrix}
	\transition{\bezpoly}{\bezpoly}^q = \transition{\lagpoly}{\bezpoly} \transition{\bezpoly}{\lagpoly}^q,
\end{equation}
which allows us to go directly from the Bézier coefficients on the domain to the Bézier coefficients on each subdomain.

\paragraph{Time subdivision matrices}
Time subdivision matrices defined by plugging the time subdivision maps 
\begin{equation}
	\psi^-(\xi,t) = (\xi, t/2), \qquad \psi^+(\xi,t) = (\xi, (t+1)/2)
\end{equation}
in \cref{eq:q-BtoL} and using \cref{eq:subdiv-matrix} as above.

\paragraph{Representations of the Jacobian determinant}

Given a standard element $\sigma_s^n$  as above, let us 
define $m=np-s$ for simplicial elements and $m=np-1$ for tensor product and mixed elements.
Let %
us denote $i_0=m-(\sum_{j=1}^{s}i_j)$ and $\xi_0=1-(\sum_{j=1}^{s}\xi_j)$. 

The Lagrange basis polynomials used to represent the Jacobian determinant $|J_x(\xi)|$ on reference element $\sigma=\sigma^n_s$ are:
\begin{gather}
	\lagpoly_{i_1,\dots,i_n}^\sigma(\xi_1,\dots,\xi_n) =
		\left(\prod_{j=0}^{s} \ell^{i_j,m}_{i_j}(\xi_{i_j})\right)
		\left(\prod_{j=s+1}^{n} \ell^{m,m}_{i_j}(\xi_{i_j})\right)
	\\
	\ell^{q,m}_{j}(\zeta) = \prod_{k\in\{0,\dots,q\}\setminus \{j\}}\frac{m\zeta - k}{j - k}
\end{gather}

The Bézier basis polynomials used to represent the Jacobian determinant $|J_x(\xi)|$ on reference element $\sigma=\sigma^n_s$ are:
\begin{gather}
	\bezpoly_{i_1,\dots,i_n}^{\sigma}(\xi_1,\dots,\xi_n) =
		\left(\binom{m}{i_0,\dots,i_s} \xi_0^{i_0} \dots \xi_s^{i_s} \right)
		\prod_{j=s+1}^{n} b^{m}_{i_j}(\xi_{i_j})
	\\
	b^q_j(\zeta) = \binom{q}{j} \zeta^{j}(1-\zeta)^{q-j}
\end{gather}
where
$\bezpoly_{i_1,\dots,i_n}^{\sigma}$ is the product of
an order $m$ Bernstein polynomial on the $s$-simplex basis, in the variables $\xi_1,\dots,\xi_s$,
with an order $m$ Bernstein polynomial on the $(n-s)$-tensor product basis in the variables $\xi_{s+1},\dots,\xi_n$, which is itself a product of $(n-s)$ univariate Bernstein polynomials of order $m$.
It is easy to see that
\begin{itemize}
	\item the order of $|J_{x}(\xi)|$ in $\{\xi_1,\dots,\xi_s\}$ is $np-s$;
	\item the order of $|J_{x}(\xi)|$ in $\{\xi_{s+1},\dots,\xi_n\}$ is $np-1$.
\end{itemize}
Notice that on $n$-simplices ($s=n$) and $n$-hypercubes ($s=1$), $|J_x|$ has the same order in every variable.

\paragraph*{Extension to Dynamic} For the dynamic case, we add a term for time to the Bézier basis construction, an order $n$ univariate Bernstein polynomial in $t$:
\begin{gather}
	\label{eq:timedep-basis}
	\timedep{\bezpoly}_{i_1,\dots,i_n}^{\sigma}(\xi_1,\dots,\xi_n,t) =
		\bezpoly_{i_1,\dots,i_n,i_t}^{\sigma}(\xi_1,\dots,\xi_n)
		b^n_{i_t}(t)
\end{gather}

It is easy to see that the order of $|J_{\timedep{x}}|$ remains the same as the static case in its spatial variables, while it is $n$ in its time variable. 
The number of polynomials in the Bernstein basis, which is equal to the number of terms of the given polynomial, increases combinatorially with the dimension $n$ and order $p$ and is given by 
\begin{equation}
	N = \binom{np}{s} (np)^{n-s}, \qquad \timedep{N} = N(n+1),
\end{equation}
for the static and the dynamic case, respectively.
Note that, this combinatorial growth poses intrinsic limitations to scaling the problem up in degree and dimension. Nevertheless, it remains tractable for the most common cases, as exemplified in \cref{tab:degrees}.

\begin{table}
	\begin{tabular}{l|cc|c|ccc|r}
		Element name & $n$ & $s$ & $p$ & $\xi_{1\dots s}$ & $\xi_{s+1 \dots n}$ & $t$ & $\timedep{N}$ \\
		\hline
		Linear triangle & 2 & 2 & 1 & 0 & - & 2 & 3 \\
		Quadratic triangle & 2 & 2 & 2 & 2 & - & 2 & 18 \\
		Cubic triangle & 2 & 2 & 3 & 4 & - & 2 & 45\\
		Quartic triangle & 2 & 2 & 4 & 6 & - & 2 & 84\\
		Quintic triangle & 2 & 2 & 5 & 8 & - & 2 & 135\\
		Bilinear quadrangle & 2 & 1 & 1 & 1 & 1 & 2 & 12\\
		Biquadratic quadrangle & 2 & 1 & 2 & 3 & 3 & 2 & 48\\
		Bicubic quadrangle & 2 & 1 & 3 & 5 & 5 & 2 & 108\\
		Linear tetrahedron & 3 & 3 & 1 & 0 & - & 3 & 4\\
		Quadratic tetrahedron & 3 & 3 & 2 & 3 & - & 3 & 80\\
		Cubic tetrahedron & 3 & 3 & 3 & 6 & - & 3 & 256\\
		Quartic tetrahedron & 3 & 3 & 4 & 9 & - & 3 & 880\\
		Bilinear tri. prism & 3 & 2 & 1 & 1 & 2 & 3 & 36\\
		Biquadratic tri. prism & 3 & 2 & 2 & 4 & 5 & 3 & 270\\
		Trilinear hexahedron & 3 & 1 & 1 & 2 & 2 & 3 & 108\\
		Triquadratic hexahedron & 3 & 1 & 2 & 5 & 5 & 3 & 864\\
	\end{tabular}
	\caption{Order of $|J_{\timedep{x}}(\xi)|$ in the spatial variables $\xi$ and the time variable $t$, and the number $\timedep{N}$ of its terms for a generic order $p$ map on $\timedep{\sigma}^n_s$. 
 }
	\label{tab:degrees}
\end{table}

\section{Subdivision rules}
\label{app:subdivision}
The subdivision functions $\psi^q$ for the various types of elements are listed below. 
Their related subdomains are shown in the insets (red bullet = origin).
The dynamic elements for $n=3$ require 4D hypercubes (for tensor product) and hyper-prisms (for simplicial and mixed).

\paragraph{Triangle}
~

\begin{minipage}{0.7\linewidth}
\small{
\begin{align*}
&\psi^1(\xi) = (\xi_1/2,\xi_2/2) \hspace*{2cm}\\
&\psi^2(\xi) = ((\xi_1+1)/2,\xi_2/2)\\
&\psi^3(\xi) = (\xi_1/2,(\xi_2+1)/2)\\
&\psi^4(\xi) = ((1-\xi_1)/2,(1-\xi_2)/2)
\end{align*}
}
\end{minipage}%
\begin{minipage}{0.3\linewidth}
\hspace*{-0.5cm}
\tdplotsetmaincoords{0}{0}
\begin{tikzpicture}[scale=1.8, tdplot_main_coords]	
        \draw node at (0.16,0.16) {1};
        \draw node at (0.67,0.16) {2};
        \draw node at (0.16,0.67) {3};
        \draw node at (0.33,0.33) {4};
        
	\draw (0,0) -- (0,1) -- (1,0) -- cycle
        (0.5,0) -- (0.5,0.5) -- (0,0.5) -- cycle;
	
	\draw[radius=1pt, fill=red] (0,0) circle;
	\draw[radius=1pt, fill=blue] (1,0) circle;
	\draw[radius=1pt, fill=blue] (0,1) circle;
	\draw[radius=1pt, fill=blue] (0.5,0) circle;
	\draw[radius=1pt, fill=blue] (0.5,0.5) circle;
	\draw[radius=1pt, fill=blue] (0,0.5) circle;
\end{tikzpicture}
\end{minipage}

\paragraph{Quad}
~

\begin{minipage}{0.7\linewidth}
\small{
\begin{align*}
&\psi^1(\xi) = (\xi_1/2,\xi_2/2) \hspace*{2cm}\\
&\psi^2(\xi) = ((\xi_1+1)/2,\xi_2/2)\\
&\psi^3(\xi) = (\xi_1/2,(\xi_2+1)/2)\\
&\psi^4(\xi) = ((\xi_1+1)/2,(\xi_2+1)/2)
\end{align*}
}
\end{minipage}%
\begin{minipage}{0.3\linewidth}
  \hspace*{-0.5cm}
\tdplotsetmaincoords{0}{0}
\begin{tikzpicture}[scale=1.8, tdplot_main_coords]	
        \draw node at (0.25,0.25) {1};
        \draw node at (0.75,0.25) {2};
        \draw node at (0.25,0.75) {3};
        \draw node at (0.75,0.75) {4};
	\draw (0,0) -- (0,1) -- (1,1) -- (1,0) -- cycle
            (0,0) -- (0.5,0) -- (0.5,0.5) -- (0,0.5) -- cycle
            (0.5,0) -- (1,0) -- (1,0.5) -- (0.5,0.5) -- cycle
            (0,0.5) -- (0.5,0.5) -- (0.5,1) -- (0,1) -- cycle
            (0.5,0.5) -- (0.5,1) -- (1,1) -- (0.5,1) -- cycle;
	
	\draw[radius=1pt, fill=red] (0,0) circle;
	\draw[radius=1pt, fill=blue] (1,0) circle;
	\draw[radius=1pt, fill=blue] (0,1) circle;
	\draw[radius=1pt, fill=blue] (1,1) circle;
	\draw[radius=1pt, fill=blue] (0.5,0) circle;
	\draw[radius=1pt, fill=blue] (0.5,0.5) circle;
	\draw[radius=1pt, fill=blue] (0,0.5) circle;
	\draw[radius=1pt, fill=blue] (1,0.5) circle;
	\draw[radius=1pt, fill=blue] (0.5,1) circle;
\end{tikzpicture}
\end{minipage}%

\paragraph{Tetrahedron}
The subdivision of the tetrahedron is non-trivial: 
it is split into four tetrahedra incident at the corners of the domain (corresponding to subdomains 1--4) and a central octahedron, which is further split into four tetrahedra (domains 5--8).

\begin{minipage}{0.8\linewidth}
\small{
\begin{align*}
&\psi^1(\xi) = (\xi_1/2,\xi_2/2,\xi_3/2) \hspace*{5cm}\\
&\psi^2(\xi) = ((\xi_1+1)/2,\xi_2/2,\xi_3/2)\\
&\psi^3(\xi) = (\xi_1/2,(\xi_2+1)/2,\xi_3/2)\\
&\psi^4(\xi) = (\xi_1/2,\xi_2/2,(\xi_3+1)/2)\\
&\psi^5(\xi) = ((1-\xi_2-\xi_3)/2,\xi_2/2,(\xi_1+\xi_2+\xi_3)/2)\\
&\psi^6(\xi) = ((1-\xi_2)/2,(\xi_1+\xi_2)/2,(\xi_2+\xi_3)/2)\\
&\psi^7(\xi) = ((\xi_1+\xi_2)/2,(1-\xi_1)/2,\xi_3/2)\\
&\psi^8(\xi) = (\xi_1/2,(\xi_2+\xi_3)/2,(1-\xi_1+\xi_2)/2)\\
\end{align*}
}
\end{minipage}%
\hspace*{-2.3cm}
\begin{minipage}{0.4\linewidth}
\tdplotsetmaincoords{50}{120}
\begin{tikzpicture}[scale=1.6, tdplot_main_coords]	
	\draw (0,0,1) -- (0,1,0) -- (1,0,0) -- cycle 
		(0,0,1) -- (1,0,0) -- (0,0,0) -- cycle 
		(0,1,0) -- (0,0,1) -- (0,0,0) -- cycle 
		(1,0,0) -- (0,1,0) -- (0,0,0) -- cycle 
		(0,0,1) -- (0,0,0)
		(0,1,0) -- (1,0,0) -- (0,0,0) -- (0,1,0) -- (0,0,1)
		(1,0,0) -- (0,0,1);
	
	\draw[radius=1pt, fill=red] (0,0,0) circle;
	\draw[radius=1pt, fill=blue] (1,0,0) circle;
	\draw[radius=1pt, fill=blue] (0,1,0) circle;
	\draw[radius=1pt, fill=blue] (0,0,1) circle;

         \draw (0.5,0.5,0) -- (0,0.5,0) -- (0.5,0,0) -- cycle
                (0.5,0,0) -- (0.5,0,0.5) -- (0,0,0.5) -- cycle
                (0,0.5,0) -- (0,0.5,0.5) -- (0,0,0.5) -- cycle
                (0.5,0.5,0) -- (0.5,0,0.5) -- (0,0.5,0.5) -- cycle
                (0,0,0.5) -- (0.5,0.5,0);

        \draw[radius=1pt, fill=blue] (0.5,0,0) circle;
        \draw[radius=1pt, fill=blue] (0.5,0.5,0) circle;
        \draw[radius=1pt, fill=blue] (0,0.5,0) circle;
        \draw[radius=1pt, fill=blue] (0.5,0,0.5) circle;
        \draw[radius=1pt, fill=blue] (0,0,0.5) circle;
        \draw[radius=1pt, fill=blue] (0,0.5,0.5) circle;

\end{tikzpicture}
\end{minipage}

\paragraph{Hexa}
The subdivision of a hexahedron works both for a static hexahedral domain and a dynamic quad domain. 
In the latter case, the third coordinate is time.

\begin{minipage}{0.6\linewidth}
\small{
\begin{align*}
&\psi^1(\xi) = (\xi_1/2,\xi_2/2,\xi_3/2) \hspace*{4.5cm}\\
&\psi^2(\xi) = ((\xi_1+1)/2,\xi_2/2,\xi_3/2)\\
&\psi^3(\xi) = (\xi_1/2,(\xi_2+1)/2,\xi_3/2)\\
&\psi^4(\xi) = ((\xi_1+1)/2,(\xi_2+1)/2,\xi_3/2)\\
&\psi^5(\xi) = (\xi_1/2,\xi_2/2,(\xi_3+1)/2)\\
&\psi^6(\xi) = ((\xi_1+1)/2,\xi_2/2,(\xi_3+1)/2)\\
&\psi^7(\xi) = (\xi_1/2,(\xi_2+1)/2,(\xi_3+1)/2)\\
&\psi^8(\xi) = ((\xi_1+1)/2,(\xi_2+1)/2,(\xi_3+1)/2)
\end{align*}
}
\end{minipage}%
\begin{minipage}{0.2\linewidth}
\hspace*{-2.5cm}
\tdplotsetmaincoords{70}{120}
\begin{tikzpicture}[scale=1.8, tdplot_main_coords]	
	\draw (0,0,0) -- (1,0,0) -- (1,1,0) -- (0,1,0) -- cycle 
		(0,0,1) -- (1,0,1) -- (1,1,1) -- (0,1,1) -- cycle 
	    (0,0,0) -- (1,0,0) -- (1,0,1) -- (0,0,1) -- cycle 
	    (0,1,0) -- (1,1,0) -- (1,1,1) -- (0,1,1) -- cycle 
		(1,0,0) -- (1,1,0) -- (1,1,1) -- (1,0,1) -- cycle 
		(0,0,0) -- (0,1,0) -- (0,1,1) -- (0,0,1) -- cycle;
  
	\draw[radius=1pt, fill=red] (0,0,0) circle;
	\draw[radius=1pt, fill=blue] (1,0,0) circle;
	\draw[radius=1pt, fill=blue] (0,1,0) circle;
	\draw[radius=1pt, fill=blue] (1,1,0) circle;
	\draw[radius=1pt, fill=blue] (0,0,1) circle;
	\draw[radius=1pt, fill=blue] (1,0,1) circle;
	\draw[radius=1pt, fill=blue] (0,1,1) circle;
	\draw[radius=1pt, fill=blue] (1,1,1) circle;

    \draw (0,0,0.5) -- (1,0,0.5) -- (1,1,0.5) -- (0,1,0.5) -- cycle
	    (0,0.5,0) -- (1,0.5,0) -- (1,0.5,1) -- (0,0.5,1) -- cycle 
		(0.5,0,0) -- (0.5,1,0) -- (0.5,1,1) -- (0.5,0,1) -- cycle
            (0.5,0.5,0) -- (0.5,0.5,1)
            (0.5,0,0.5) -- (0.5,1,0.5)
            (0,0.5,0.5) -- (1,0.5,0.5);

  	\draw[radius=1pt, fill=blue] (0.5,0,0) circle;
  	\draw[radius=1pt, fill=blue] (0.5,0.5,0) circle;
  	\draw[radius=1pt, fill=blue] (0.5,0.5,0.5) circle;
  	\draw[radius=1pt, fill=blue] (0,0.5,0) circle;
  	\draw[radius=1pt, fill=blue] (0,0.5,0.5) circle;
  	\draw[radius=1pt, fill=blue] (0,0,0.5) circle;
  	\draw[radius=1pt, fill=blue] (0.5,0,0.5) circle;
  	\draw[radius=1pt, fill=blue] (0.5,1,0) circle;
  	\draw[radius=1pt, fill=blue] (0.5,0,1) circle;
  	\draw[radius=1pt, fill=blue] (0.5,0.5,1) circle;
  	\draw[radius=1pt, fill=blue] (1,0.5,0) circle;
  	\draw[radius=1pt, fill=blue] (0,0.5,1) circle;
  	\draw[radius=1pt, fill=blue] (1,0.5,0.5) circle;
  	\draw[radius=1pt, fill=blue] (1,0,0.5) circle;
  	\draw[radius=1pt, fill=blue] (0,1,0.5) circle;
  	\draw[radius=1pt, fill=blue] (0.5,1,0.5) circle;
  	\draw[radius=1pt, fill=blue] (1,1,0.5) circle;
  	\draw[radius=1pt, fill=blue] (1,0.5,1) circle;
  	\draw[radius=1pt, fill=blue] (0.5,1,1) circle;
\end{tikzpicture}
\end{minipage}

\paragraph{Prism}
The subdivision of a prism works both for a static prism domain and a dynamic triangle domain. 
In the latter case, the third coordinate is time.

\begin{minipage}{0.8\linewidth}
\small{
\begin{align*}
&\psi^1(\xi) = (\xi_1/2,\xi_2/2,\xi_3/2) \hspace*{4.5cm}\\
&\psi^2(\xi) = ((\xi_1+1)/2,\xi_2/2,\xi_3/2)\\
&\psi^3(\xi) = (\xi_1/2,(\xi_2+1)/2,\xi_3/2)\\
&\psi^4(\xi) = ((1-\xi_1)/2,(1-\xi_2)/2,\xi_3/2)\\
&\psi^5(\xi) = (\xi_1/2,\xi_2/2,(\xi_3+1)/2)\\
&\psi^6(\xi) = ((\xi_1+1)/2,\xi_2/2,(\xi_3+1)/2)\\
&\psi^7(\xi) = (\xi_1/2,(\xi_2+1)/2,(\xi_3+1)/2)\\
&\psi^8(\xi) = ((1-\xi_1)/2,(1-\xi_2)/2,(\xi_3+1)/2)\\
\end{align*}
}
\end{minipage}%
\begin{minipage}{0.4\linewidth}
  \hspace{-2.5cm}
\tdplotsetmaincoords{70}{120}
\begin{tikzpicture}[scale=1.6, tdplot_main_coords]	
	\draw (0,0,0) -- (1,0,0) -- (0,1,0) -- cycle 
		(0,0,1) -- (1,0,1) -- (0,1,1) -- cycle 
		(0,0,0) -- (0,0,1)  
		(1,0,0) -- (1,0,1)
		(0,1,0) -- (0,1,1);
	
	\draw[radius=1pt, fill=red] (0,0,0) circle;
	\draw[radius=1pt, fill=blue] (1,0,0) circle;
	\draw[radius=1pt, fill=blue] (0,1,0) circle;
	\draw[radius=1pt, fill=blue] (0,0,1) circle;
	\draw[radius=1pt, fill=blue] (1,0,1) circle;
	\draw[radius=1pt, fill=blue] (0,1,1) circle;

        \draw (0.5,0.5,0) -- (0,0.5,0) -- (0.5,0,0) -- cycle
                (0.5,0.5,1) -- (0,0.5,1) -- (0.5,0,1) -- cycle
                (0.5,0.5,0.5) -- (0,0.5,0.5) -- (0.5,0,0.5) -- cycle
                (0.5,0.5,0) -- (0.5,0.5,1)
                (0,0.5,0) -- (0,0.5,1)
                (0.5,0,0) -- (0.5,0,1)
                (0,0,0.5) -- (1,0,0.5) -- (0,1,0.5) -- cycle;
                
	\draw[radius=1pt, fill=blue] (0,0,0.5) circle;
	\draw[radius=1pt, fill=blue] (1,0,0.5) circle;
	\draw[radius=1pt, fill=blue] (0,1,0.5) circle;
	\draw[radius=1pt, fill=blue] (0.5,0.5,0) circle;
	\draw[radius=1pt, fill=blue] (0,0.5,0) circle;
	\draw[radius=1pt, fill=blue] (0.5,0,0) circle;
	\draw[radius=1pt, fill=blue] (0.5,0.5,1) circle;
	\draw[radius=1pt, fill=blue] (0,0.5,1) circle;
	\draw[radius=1pt, fill=blue] (0.5,0,1) circle;
	\draw[radius=1pt, fill=blue] (0.5,0.5,0.5) circle;
	\draw[radius=1pt, fill=blue] (0,0.5,0.5) circle;
	\draw[radius=1pt, fill=blue] (0.5,0,0.5) circle;                
\end{tikzpicture}
\end{minipage}

\paragraph{Hypercube}
This is a 4-hypercube that works as a domain for a dynamic hexahedral element.
\small{
\begin{align*}
&\psi^1(\xi,t) = (\xi_1/2,\xi_2/2,\xi_3/2,t/2,t/2) \hspace*{3.3cm}\\
&\psi^2(\xi,t) = ((\xi_1+1)/2,\xi_2/2,\xi_3/2,t/2,t/2)\\
&\psi^3(\xi,t) = (\xi_1/2,(\xi_2+1)/2,\xi_3/2,t/2)\\
&\psi^4(\xi,t) = ((\xi_1+1)/2,(\xi_2+1)/2,\xi_3/2,t/2)\\
&\psi^5(\xi,t) = (\xi_1/2,\xi_2/2,(\xi_3+1)/2,t/2)\\
&\psi^6(\xi,t) = ((\xi_1+1)/2,\xi_2/2,(\xi_3+1)/2,t/2)\\
&\psi^7(\xi,t) = (\xi_1/2,(\xi_2+1)/2,(\xi_3+1)/2,t/2)\\
&\psi^8(\xi,t) = ((\xi_1+1)/2,(\xi_2+1)/2,(\xi_3+1)/2,t/2)\\
&\psi^9(\xi,t) = (\xi_1/2,\xi_2/2,\xi_3/2,t/2,(t+1)/2)\\
&\psi^{10}(\xi,t) = ((\xi_1+1)/2,\xi_2/2,\xi_3/2,t/2,(t+1)/2)\\
&\psi^{11}(\xi,t) = (\xi_1/2,(\xi_2+1)/2,\xi_3/2,(t+1)/2)\\
&\psi^{12}(\xi,t) = ((\xi_1+1)/2,(\xi_2+1)/2,\xi_3/2,(t+1)/2)\\
&\psi^{13}(\xi,t) = (\xi_1/2,\xi_2/2,(\xi_3+1)/2,(t+1)/2)\\
&\psi^{14}(\xi,t) = ((\xi_1+1)/2,\xi_2/2,(\xi_3+1)/2,(t+1)/2)\\
&\psi^{15}(\xi,t) = (\xi_1/2,(\xi_2+1)/2,(\xi_3+1)/2,(t+1)/2)\\
&\psi^{16}(\xi,t) = ((\xi_1+1)/2,(\xi_2+1)/2,(\xi_3+1)/2,(t+1)/2)
\end{align*}
}

\paragraph{Hyperprism 1}
This is the tensor product of a tetrahedron with an interval that works as a domain for a dynamic tetrahedral element.

\small{
\begin{align*}
&\psi^1(\xi,t) = (\xi_1/2,\xi_2/2,\xi_3/2,t/2) \hspace*{4cm}\\
&\psi^2(\xi,t) = ((\xi_1+1)/2,\xi_2/2,\xi_3/2,t/2)\\
&\psi^3(\xi,t) = (\xi_1/2,(\xi_2+1)/2,\xi_3/2,t/2)\\
&\psi^4(\xi,t) = (\xi_1/2,\xi_2/2,(\xi_3+1)/2,t/2)\\
&\psi^5(\xi,t) = ((1-\xi_2-\xi_3)/2,\xi_2/2,(\xi_1+\xi_2+\xi_3)/2,t/2)\\
&\psi^6(\xi,t) = ((1-\xi_2)/2,(\xi_1+\xi_2)/2,(\xi_2+\xi_3)/2,t/2)\\
&\psi^7(\xi,t) = ((\xi_1+\xi_2)/2,(1-\xi_1)/2,\xi_3/2,t/2)\\
&\psi^8(\xi,t) = (\xi_1/2,(\xi_2+\xi_3)/2,(1-\xi_1+\xi_2)/2,t/2)\\
&\psi^9(\xi,t) = (\xi_1/2,\xi_2/2,\xi_3/2,(t+1)/2)\\
&\psi^{10}(\xi,t) = ((\xi_1+1)/2,\xi_2/2,\xi_3/2,(t+1)/2)\\
&\psi^{11}(\xi,t) = (\xi_1/2,(\xi_2+1)/2,\xi_3/2,(t+1)/2)\\
&\psi^{12}(\xi,t) = (\xi_1/2,\xi_2/2,(\xi_3+1)/2,(t+1)/2)\\
&\psi^{13}(\xi,t) = ((1-\xi_2-\xi_3)/2,\xi_2/2,(\xi_1+\xi_2+\xi_3)/2,(t+1)/2)\\
&\psi^{14}(\xi,t) = ((1-\xi_2)/2,(\xi_1+\xi_2)/2,(\xi_2+\xi_3)/2,(t+1)/2)\\
&\psi^{15}(\xi,t) = ((\xi_1+\xi_2)/2,(1-\xi_1)/2,\xi_3/2,(t+1)/2)\\
&\psi^{16}(\xi,t) = (\xi_1/2,(\xi_2+\xi_3)/2,(1-\xi_1+\xi_2)/2,(t+1)/2)
\end{align*}
}

\paragraph{Hyperprism 2}
This is the tensor product of a prism with an interval that works as a domain for a dynamic prism element. 
\small{
\begin{align*}
&\psi^1(\xi.t) = (\xi_1/2,\xi_2/2,\xi_3/2,t/2) \hspace*{4cm}\\
&\psi^2(\xi.t) = ((\xi_1+1)/2,\xi_2/2,\xi_3/2,t/2)\\
&\psi^3(\xi.t) = (\xi_1/2,(\xi_2+1)/2,\xi_3/2,t/2)\\
&\psi^4(\xi.t) = ((1-\xi_1)/2,(1-\xi_2)/2,\xi_3/2,t/2)\\
&\psi^5(\xi.t) = (\xi_1/2,\xi_2/2,(\xi_3+1)/2,t/2)\\
&\psi^6(\xi.t) = ((\xi_1+1)/2,\xi_2/2,(\xi_3+1)/2,t/2)\\
&\psi^7(\xi.t) = (\xi_1/2,(\xi_2+1)/2,(\xi_3+1)/2,t/2)\\
&\psi^8(\xi.t) = ((1-\xi_1)/2,(1-\xi_2)/2,(\xi_3+1)/2,t/2)\\
&\psi^9(\xi.t) = (\xi_1/2,\xi_2/2,\xi_3/2,(t+1)/2)\\
&\psi^{10}(\xi.t) = ((\xi_1+1)/2,\xi_2/2,\xi_3/2,(t+1)/2)\\
&\psi^{11}(\xi.t) = (\xi_1/2,(\xi_2+1)/2,\xi_3/2,(t+1)/2)\\
&\psi^{12}(\xi.t) = ((1-\xi_1)/2,(1-\xi_2)/2,\xi_3/2,(t+1)/2)\\
&\psi^{13}(\xi.t) = (\xi_1/2,\xi_2/2,(\xi_3+1)/2,(t+1)/2)\\
&\psi^{14}(\xi.t) = ((\xi_1+1)/2,\xi_2/2,(\xi_3+1)/2,(t+1)/2)\\
&\psi^{15}(\xi.t) = (\xi_1/2,(\xi_2+1)/2,(\xi_3+1)/2,(t+1)/2)\\
&\psi^{16}(\xi.t) = ((1-\xi_1)/2,(1-\xi_2)/2,(\xi_3+1)/2,(t+1)/2)
\end{align*}
}

\section{Interval arithmetic}
\label{app:interval}
The interval type we use defines the following operations and relations:
\begin{gather*}
	-[\intlo{x}, \inthi{x}] = [-\inthi{x}, -\intlo{x}]
	\\
	[\intlo{x}, \inthi{x}] + [\intlo{y}, \inthi{y}] =
	[\intlo{x} + \intlo{y}, \inthi{x} + \inthi{y}]
	\\
	[\intlo{x}, \inthi{x}] [\intlo{y}, \inthi{y}] =
	[\min(\intlo{x}\intlo{y},\intlo{x}\inthi{y},\inthi{x}\intlo{y},\inthi{x}\inthi{y}),
	\max(\intlo{x}\intlo{y},\intlo{x}\inthi{y},\inthi{x}\intlo{y},\inthi{x}\inthi{y})]
	\\
	\min([\intlo{x}, \inthi{x}], [\intlo{y}, \inthi{y}]) = [\min(\intlo{x}, \intlo{y}), \min(\inthi{x}, \inthi{y})]
	\\
	\max([\intlo{x}, \inthi{x}], [\intlo{y}, \inthi{y}]) = [\max(\intlo{x}, \intlo{y}), \max(\inthi{x}, \inthi{y})]
	\\
	[\intlo{x}, \inthi{x}] = [\intlo{y}, \inthi{y}] \Leftrightarrow \intlo{x} = \intlo{y} \wedge \inthi{x} = \inthi{y}
	\\
	[\intlo{x}, \inthi{x}] > [\intlo{y}, \inthi{y}] \Leftrightarrow \intlo{x} > \inthi{y}
\end{gather*}

The division between intervals should be avoided, as the divisor may contain 0.
However, a division of an interval by an exact number is acceptable; in our code, we have only divisions by 2, which is however exact even in floating point.

Note that there is no total ordering of intervals, so for example, $[\intlo{x},\inthi{x}]\nleq 0\nRightarrow [\intlo{x},\inthi{x}]>0$.

In our implementation, outward rounding is achieved by internally storing the lower end of the interval with the opposite sign and changing the processor rounding mode to always rounding up. This ensures that the interval's width increases only when the actual floating-point computation is inexact. 

\section{Ground truth}
\label{app:mathematica}
We designed a Mathematica script that computes symbolically the roots of $|J_{\timedep x}|$ for a given %
element. 
The script receives in input the type, dimension $n$, and order $p$ of an element together with the control points describing its geometric map at times $t=0$ and $t=1$.
This is the same input taken by our test.
The script computes the symbolic expression of $|J_{\timedep x}|$, which is a polynomial in $n+1$ variables, finds its roots inside the domain $\timedep{\sigma}$, %
and returns the minimum value of $t$ in which a root is found.
First, we use \texttt{FindInstance} to check if the region where $J\leq 0$ is empty. 
If so, the element is valid throughout; otherwise, we use \texttt{Minimize} to find the minimum time $t$ in such a region. 
The scripts will be released as part of our open-source code.

\section{Failure case for \protect\cite{johnen_geometrical_2014}}
\label{app:gmsh}
We implemented the method described in \cite{johnen_geometrical_2014} for static validity with floating point arithmetic and constructed a failure case for the following invalid quadratic triangle (when tested with Mathematica).
Control points $\mathbf{v}_{00}$, $\mathbf{v}_{20}$, and $\mathbf{v}_{02}$ are the three vertices of the triangle, and $\mathbf{v}_{01}$, $\mathbf{v}_{11}$, $\mathbf{v}_{10}$ are the edge midpoints.
\begin{gather*}
    \mathbf{v}_{00} = \left( 0.33333333333333331, 0.33333333333333331 \right) \\
    \mathbf{v}_{20} = \left( 1.3333333333333333, 0.33333333333333331 \right) \\
    \mathbf{v}_{02} = \left( 0.33333333333333331, 1.3333333333333333 \right) \\
    \mathbf{v}_{01} = \left( 0.83333333333333326, 0.33333333333333331 \right) \\
    \mathbf{v}_{11} = \left( 0.83333333333333326, 0.58333333333333315 \right) \\
    \mathbf{v}_{10} = \left( 0.33333333333333331, 0.83333333333333326 \right)
\end{gather*}

\end{document}